\documentclass{emulateapj}
\usepackage{booktabs}

\newcommand{\diff}{\mathrm{d}}
\defcitealias{BL}{BL09}

\slugcomment{}

\shorttitle{Symmetrically Limb-brightened Jets}
\shortauthors{Takahashi et al.}

\begin{document}

\title{Fast-spinning black holes inferred from symmetrically limb-brightened radio jets}

\author{Kazuya Takahashi\altaffilmark{1}, Kenji Toma\altaffilmark{2,3}, Motoki Kino\altaffilmark{4,5}, Masanori Nakamura\altaffilmark{6} and Kazuhiro Hada\altaffilmark{5}}
\affil{$^1$ Center for Gravitational Physics, Yukawa Institute for Theoretical Physics, Kyoto University, Kyoto, 606-8502, Japan}

\altaffiltext{2}{Frontier Research Institute for Interdisciplinary Sciences, Tohoku University, Sendai, 980-8578, Japan}
\altaffiltext{3}{Astronomical Institute, Tohoku University, Sendai, 980-8578, Japan}
\altaffiltext{4}{Kogakuin University, 2665-1, Nakano-cho, Hachioji-chi, Tokyo, 192-0015, Japan}
\altaffiltext{5}{National Astronomical Observatory of Japan, Mitaka, Tokyo 181-8588, Japan}
\altaffiltext{6}{Institute of Astronomy \& Astrophysics, Academia Sinica, P.O. Box 23-141, Taipei 10617, Taiwan}

\begin{abstract}
This paper theoretically investigates the relations between the structure of relativistic jets and produced synchrotron images, by using a steady axisymmetric force-free jet model. We especially focus on the limb-brightened jets that are largely symmetric to the jet axes and observed in some active galactic nuclei such as M87, Mrk~501, Cyg~A, and 3C84. We find that symmetrically limb-brightened images can be produced when magnetic field lines of the jet penetrate a fast-spinning black hole as motivated by the Blandford-Znajek mechanism. On the other hand, jets with magnetic field lines that pass through a slowly spinning black hole or the Keplerian accretion disk produce highly asymmetric radio images. In addition, the edge of a counter jet tends to be luminous in the accretion-disk model even for rather small viewing angles, which may be problematic for some observed jets. We also suggest that the site of particle accelerations in relativistic jets can be constrained by fitting the radio images to observations. This kind of study focusing on the jet images far away from the central engine is complimentary to those concentrating directly on the innermost region with upcoming data of the Event Horizon Telescope.
\end{abstract}


\section{Introduction}
The launching mechanism of collimated relativistic outflows (jets) is one of the mysteries in astrophysics, which are observed in active galactic nuclei (AGNs) and micro-quasars, and most probably associated with gamma-ray bursts and some tidal disruption events. They are believed to be launched from a system with a black hole (BH) and the accretion disk. In particular, AGN jets are widely thought to be electromagnetically launched along globally ordered magnetic field lines from the BHs via the Blandford-Znajek (BZ) mechanism \citep{BZ} and/or from the accretion disks via unipolar induction mechanism \citep{BP}. While the BZ mechanism effectively works to drive a Poynting-flux-dominated jet in general relativistic magnetohydrodynamics (GRMHD) simulations \citep{MG04,Komissarov05,BK08,MB09,TNM11,Qian18} and in those with radiation \citep[GRRMHD simulations;][]{McKinney14,Sadowski14}, the real footpoint of astrophysical jets is yet to be confirmed from observations. Observational evidence for the BZ mechanism, if any, would support the existence of ergosphere \citep{Komissarov04,TT14,TT16,KI17}.

Radio observations with VLBI techniques now come to resolve a jet at the very vicinity of the central BH \citep{Junor99}. Recently, \citet{Hada16} revealed an evident limb-brightened feature of the jet of M87 at $\sim 0.5$~mas from the BH, which corresponds to $140\mathrm{-}280$ Schwarzschild radii ($r_g=2GM_\mathrm{BH}/c^2$) for the distance to M87 \citep[$D=16.7$~Mpc;][]{D} and the BH mass \citep[$M_{\rm BH} \sim (3 \mathrm{-} 6) \times 10^9 M_\odot$;][]{M6,M3}. The limb-brightened feature is largely symmetric to the jet axis and observed in the downstream at least up to $\sim 10^4r_g$ (projected) from the center with VLBI \citep{Kovalev07,Walker08}, while the feature is still less clear for the faint counter jet. We note that largely symmetric limb-brightened jets are also observed in other AGNs such as Mrk~501 \citep[][and references therein]{Giroletti04}, Cyg~A \citep[e.g.][]{Boccardi16}, and 3C84 \citep{Nagai14,Giovannini18} whereas their spatial structures have been less resolved.

Theoretically, \citet[][hereafter \citetalias{BL}]{BL} proposed a steady axisymmetric jet model to synthesize radio images of the M87 jet. They supposed a paraboloid-shaped force-free magnetic field that co-rotates with a Keplerian accretion disk at the equator. Their model succeeded in reproducing a jet length similar to observations and a dim counter jet for an assumed spatial distribution of the non-thermal electrons. However, the produced images do not show limb-brightened features but illuminate the jet axis. While \citetalias{BL} focused more on the images of the BH shadow that will be detected by the Event Horizon Telescope \citep[EHT;][]{Doeleman12,Akiyama17}, it will be important to ensure the consistency of the model with the downstream observations.

In this paper, we investigate the relations between the structure of relativistic jets and observed radio images. We employ the force-free paraboloidal jet model of \citetalias{BL}, which will be suitable at least for the M87 jet, since the force-free approximation would be reasonable especially in the base of the M87 jet \citep{Kino14,Kino15} and the shape of the M87 jet can be reasonably fit by a parabola \citep{AN12,Hada13,Nakamura18}. We introduce some new physics to the model of \citetalias{BL}: Motivated by the BZ-mechanism, we newly consider jets with rigidly rotating magnetic field, as well as those with the Keplerian rotation. It is found that the difference of the jet launching point qualitatively changes the whole jet structure and leads to qualitatively and quantitatively different radio images even for the same distribution of the emitting particles. We also try more general patterns of the distribution of the non-thermal electrons, since it is not well constrained where and how particles are accelerated in relativistic jets. As shown later in this paper, we find that symmetrically limb-brightened features can be synthesized when the magnetic field lines penetrate a fast-spinning BH. Depending on the viewing angle, the counter jet becomes either luminous or dim. It is also shown, on the other hand, that symmetrically limb-brightened features cannot be produced when the magnetic field lines co-rotate with the Keplerian accretion disk, even if the non-thermal electrons are distributed on the jet edge. Since the jet model and the distribution of the non-thermal electrons are critical to produce BH shadows \citep{Dexter12,Moscibrodzka16}, this kind of study to constraint the jet base structure from the observational jet images at far zone must be complimentary to those employing the upcoming EHT data.

The paper is organized as follows: We briefly introduce our steady axisymmetric force-free model in the next section while the details are explained in Appendix~\ref{sec.model}. Section~\ref{sec.results} presents our calculated radio images for various parameter sets, where we fix some quantities to our fiducial values. The dependence on some of the fixed parameters are separately studied in Appendix~\ref{sec.dependence} while it does not affect our conclusions. We pay close attention to the difference between our force-free model and more realistic models by discussing in Sec.~\ref{sec.discussion} how our synthesized radio images can change in cold ideal MHD treatment. Effects of the viewing angle are also discussed in the latter part of Sec~\ref{sec.discussion}. We finally summarize and conclude our study in Sec.~\ref{sec.conclusion}.

\section{Method}
To simulate the radio emissions from relativistic AGN jets, we use an analytic model. The first subsection~\ref{method.jet} introduces the jet model including the magnetic and velocity fields as well as the distribution of the non-thermal electrons. The second subsection~\ref{method.image} explains the method to calculate a radio intensity map produced by synchrotron radiation. The last subsection~\ref{method.para} is devoted to the strategy to choose our model parameters.
\subsection{Force-Free Jet Model}\label{method.jet}
As shown below, our force-free model is essentially the same as in \citetalias{BL}. Although we employ a flat spacetime outside the BH, the magnetic field configuration is not much different from that with general relativistic treatment even near the hole \citep{MN07}. The resultant radio images will not be significantly changed as long as we focus on the limb-brightened features seen far from the central warped region. We put the detailed formulation in Appendix~\ref{sec.model} and briefly explain the salient results below.

\subsubsection{Electromagnetic field}
In a steady axisymmetric force-free field, a stream function $\Psi$ gives the electromagnetic field. Following \citetalias{BL}, we assume a parametrically controlled paraboloid-like-shaped jet instead of an exact solution of force-free field. The stream function is given by
\begin{equation}\label{eq.Psi}
\Psi = Ar^\nu(1\mp \cos \theta).
\end{equation}
In the above expression, $(r, \theta, \phi)$ denote the standard spherical coordinates and the minus and plus signatures are for $z\ge0$ and $z<0$, respectively. $\nu$ is the parameter to control the jet shape, where $\nu = 1$ gives paraboloidal jets, and $A$ is a constant that has the dimension of $[r^{2-\nu}B]$ with $B$ being magnetic field.
The electromagnetic field is then given by 

\begin{eqnarray}
{\bf B}_p &=&  \frac{1}{R}{\bf \nabla} \Psi \times \hat{\bf \phi}, \label{body.Bp}\\
\label{Bphitext}
B_\phi &=& \mp \frac{2\Omega_\mathrm{F} \Psi}{Rc}, \\
{\bf E} &=& -\frac{1}{c}\Omega_\mathrm{F} {\bf \nabla} \Psi = -\frac{R\Omega_\mathrm{F}}{c} \hat{\phi } \times {\bf B},\label{body.E}
\end{eqnarray}
where ${\bf B}$ and ${\bf E}$ denote the magnetic and electric fields, respectively. $(R, \phi, z)$ are the standard cylindrical coordinates and the subscript $p$ is assigned for the poloidal component. $\hat{\bf \phi}$ is the azimuthal unit vector. $\Omega_\mathrm{F}=\Omega_\mathrm{F}(\Psi)$ corresponds to the rotational frequency of magnetic fields. It should be noted that the magnetic field is wound up and toroidal-dominant, $B\sim |B_\phi|$, for $R|\Omega_{F}|/c \gg 1$ while it is poloidal-dominant, $B\sim |{\bf B}_p|$, for $R|\Omega_{F}|/c \ll 1$.

\subsubsection{Fluid velocity}\label{method.jet.v}
The fluid velocity cannot be determined in the force-free limit in principle, since the inertia is totally neglected. In this paper, we use the following drift velocity, ${\bf v}$, as fluid velocity by following \citetalias{BL}:
\begin{equation}\label{eq.drift}
{\bf v} = \frac{{\bf E}\times {\bf B}}{B^2} c = -R\Omega_\mathrm{F} \frac{B_\phi}{B^2}{\bf B}_p + R\Omega_\mathrm{F} \frac{B_p^2 }{B^2} \hat{{\bf \phi}}.
\end{equation}
The above velocity holds the following conditions for the electromagnetic field given by Eqs.~(\ref{body.Bp})-(\ref{body.E}):
(i) the velocity does not exceed the speed of light in the entire region for $\nu\le \sqrt{2}$, (ii) The electric field vanishes in the fluid rest frame (the frozen-in condition), and (iii) the velocity is asymptotically the same as in cold ideal MHD when $R|\Omega_{F}|/c \gg 1$ (See Appendix~\ref{sec.v}). Figure~\ref{fig.fieldlines} sketches an example of the twisted magnetic and velocity field lines in a paraboloidal-shaped jet ($\nu = 1$).
We note that the velocity is perpendicular to the magnetic field while their poloidal components, ${\bf v}_p$ and ${\bf B}_p$, are parallel to each other. 
The asymptotic relations of the velocity for $R|\Omega_{F}|/c \gg 1$ are given by
\begin{eqnarray}
\label{accele1}
\beta &\sim & \beta _p \sim g(\theta,\nu),\\
\label{accele2}
\beta _\phi &\sim& \left(\frac{R|\Omega_\mathrm{F}|}{c}\right)^{-1} [g(\theta,\nu)]^2, 
\end{eqnarray}
where $\beta$ denotes the speed normalized by $c$ and $g(\theta, \nu)$ is a factor that is order of tenth and approaches unity toward the jet axis ($\theta = 0, \pi$) (See Fig.~\ref{func_theta_nu}). That is, the fluid velocity is dominated by the poloidal component and becomes relativistic around the jet axis if $R|\Omega_\mathrm{F}|/c \gg 1$. For $R|\Omega_\mathrm{F}|/c \ll 1$, on the other hand, the following relations are obtained:
\begin{eqnarray}
\label{non-accele1}
\beta &\sim & \beta _\phi \sim \frac{R|\Omega_\mathrm{F}|}{c}, \\
\label{non-accele2}
\beta_p &\sim & \left(\frac{R\Omega_\mathrm{F}}{c}\right)^2 \frac{1}{g(\theta,\nu)}.
\end{eqnarray}
That is, the fluid velocity is not relativistic and dominated by the toroidal component.

\begin{figure}
	\centering \includegraphics[bb = 0 0 365 343, width = \columnwidth]{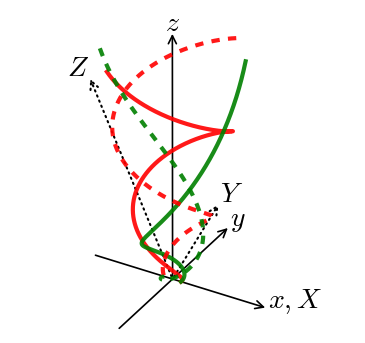}
	\caption{An example of field lines in a paraboloidal jet with $\Omega_\mathrm{F}>0$ ($z\ge 0$). The red and green lines represent magnetic field lines and stream lines, respectively. The jet axis coincides with the $z$-axis. The thick field lines originate from a point on $(x,y)$-plane while the dashed ones stem from the centrosymmetric point with respect to the origin. For visibility of the figure, we omitted the field lines in the counter jet ($z<0$), which has a symmetric structure with respect to the equatorial plane except for the direction of the poloidal magnetic field. Also plotted are $(X,Y,Z)$ coordinates, where the $X$-axis coincides with the $x$-axis and the $Z$-axis is inclined toward the $-y$ direction. The observer is assumed to be in the $Z$ direction and the angle between the $z$- and $Z$-axes corresponds to the viewing angle $\Theta$. $(X,Y)$ gives coordinates on the sky viewed from the observer.}
	\label{fig.fieldlines}
\end{figure}

\subsubsection{Non-thermal electrons}
Motivated by the limb-brightened jets, we consider the case where the non-thermal electrons are distributed away from the jet axis in contrast to \citetalias{BL}, who assumed the non-thermal electrons clinging to the axis. Such a spatial distribution concentrated away from the axis could be realized for jets launched from an accretion disk and even for those launched from the BH. The particles are supplied from a disk at the jet foot point for the former case, while several options of the particle injection can be considered for the latter case. As shown in MHD simulations \citep{MG04,Komissarov05,Komissarov07,BK08,MB09,TNM11,McKinney14,Sadowski14,Qian18}, the jets driven by the BZ mechanism are confined by the external pressure of the ambient matter, i.e., a geometrically thick disk with an advection dominated accretion flow (ADAF; \citet{NY94}) and/or the disk wind. Thermal charged particles are prevented by the globally ordered magnetic field in the funnel region from diffusing into there from the ADAF, but high-energy hadrons can diffuse into there \citep{TT12,Kimura14,Kimura15} and high-energy photons can annihilate and supply $e^-e^+$ pairs there \citep{LR11,Moscibrodzka11}. The particles in a jet flow outward from the separation surface (a.k.a.~the stagnation surface), which is much closer to the BH and the hottest part of the disk more away from the jet axis \citep{Takahashi90,MG04,Pu15,Nakamura18}.\footnote{The separation surface is the separatrix between outflowing matter that is launched as a jet and inflowing matter that is swallowed into the BH. Note that it is not taken into account in the flow velocity given by Eq.~(\ref{eq.drift}), since we only model jet outflows by neglecting general relativistic effects.} Thus, the particle injection for the outflow can be dominated at the jet edge. A pair creation gap created around the separation surface could also be a particle supplier \citep{LR11,BT15}. The fluid instability or magnetic reconnection at the layer between the jet and disk wind may also produce non-thermal particles \citep[cf.][]{MM13,Parfrey15,Toma17}.
	
It is beyond the scope of this paper to discuss in detail the above injection and acceleration mechanisms (Upcoming EHT data would shed light on those mechanisms). In this study, we simply assume that the spatial distribution of the non-thermal is described in a parametric way and the energy spectrum is given by a single power law. The spatial distribution is characterized by the cross sections at $z=\pm z_1$ for simplicity, where the electrons are assumed to be in a ring shape and the number density is given by
\begin{equation}
\label{ntext}
n(R,\pm z_1) = n_0 \exp\left[ -\frac{(R-R_p)^2}{2\Delta ^2} \right],
\end{equation}
where $R_p$ is the radius where $n$ peaks on the plane and $\Delta$ gives the width of the ring. $n_0$ is a normalization constant. Our prescription is identical to that in \citetalias{BL} when $R_p=0$ and $\Delta = z_1 = 5r_g$. We also set $z_1 = 5r_g$ hereafter while $R_p$ and $\Delta$ remain as free parameters.
In the vertical direction, $n$ is assumed to obey the continuity equation \citepalias{BL}, which is given as follows by using Eq.~(\ref{eq.drift}):\footnote{Note that \citetalias{BL} further multiplied the number density by an artificial factor of $(1 -\exp [-r^2/z_1^2])$ that works to reduce $n$ in the innermost region $r < z_1$, which is an ad-hoc treatment of gravitational effects. We do not introduce this factor while it does not change our conclusions.}
\begin{equation}\label{cnt}
\frac{n}{B^2} = \mathrm{const.} \ \mathrm{along\ a\ magnetic\ field\ line}.
\end{equation} 

We assume that the non-thermal electrons are isotropic in the fluid rest frame and obey an energy distribution of a single power law given by an index $p$:
\begin{equation}\label{fgamma}
f(\gamma') \propto \left\{
\begin{array}{cc}
\gamma' {}^{-p}& \mathrm{for}\ \gamma_\mathrm{min}' \le \gamma' \le \gamma_\mathrm{max}' \\
0& \mathrm{othewise}
\end{array},\right.
\end{equation}
where $\gamma '$ is the Lorentz factor of an electron measured in the proper frame, which have lower and higher cutoffs at $\gamma_\mathrm{min}'$ and $\gamma_\mathrm{max}'$, respectively. The synchrotron emissivity does not depend on $\gamma_\mathrm{max}'$ but only on $\gamma_\mathrm{min}'$ provided $\gamma_\mathrm{min}'$ and $\gamma_\mathrm{max}'$ are sufficiently small and large, respectively (See Appendix~\ref{sec.j}). We can, hence, set $\gamma_\mathrm{max}'=\infty$ for a large higher cutoff while we use $\gamma_\mathrm{min}'=100$ for the lower cutoff by following \citetalias{BL}. As in \citetalias{BL}, the energy distribution is fixed to Eq.~(\ref{fgamma}) everywhere, which means that some energy supplier are assumed to replenish high-energy electrons to compensate cooling processes such as the synchrotron and adiabatic coolings.

\subsection{Synchrotron Radio Images}\label{method.image}
The quantities given by Eqs.~(\ref{eq.Psi})-(\ref{cnt}) give the synchrotron emissivity at each location in a jet that is received by the observer at a frequency $\omega$ as follows \citep{RL,Shibata}:
\begin{equation}\label{jtext}
j_\omega ({\bf n}) = \frac{1}{\Gamma ^2 (1-\beta \mu)^3}j'_{\omega '}  ({\bf n}'),
\end{equation}
where the quantities with prime are evaluated in the fluid rest frame.\footnote{We excise the spherical region inside the horizon, where the emissivity is set to zero.} ${\bf n}$ is a unit vector that directs to the observer at infinity and $\mu$ is the cosine of the angle between ${\bf n}$ and ${\bf v}$. $\Gamma = 1/\sqrt{1 - \beta^2}$ is the Lorentz factor. The factor in the right hand side is attributed to relativistic effects due to the bulk fluid motion. $j'_{\omega '}({\bf n'})$ is given by Eq.~(\ref{jd}).

In higher frequencies, radio jets in AGNs are optically thin to synchrotron emissions. The intensity of radio images observed on the sky is, then, calculated by integrating Eq.~(\ref{jtext}) along the line of sight after fixing the viewing angle $\Theta$:
\begin{equation}
I _\omega (X,Y) = \int j_\omega ({\bf n}, X, Y, Z) \diff Z,
\end{equation}
where $(X,Y)$ are the coordinates of the sky and $\diff Z$ is the line element parallel to the line of sight. In the following, the $X$-axis is chosen to coincide with the $x$-axis and the $Z$-axis is inclined toward $-y$ direction so that the angle between the $z$- and $Z$-axes becomes $\Theta$ (See Fig.~\ref{fig.fieldlines}). A simulated VLBI image is obtained after the convolution with a beam kernel, which is introduced in the next sub-section.

\begin{table*}[h!t]
	\label{tab.para}
	\begin{center}
		\caption{Model Parameters}
		\begin{tabular}{lll}\hline \hline
			Quantity & Symbol & Fiducial value for Figs.~2-10\\ \hline
			Rotational frequency of the magnetic field& $\Omega_\mathrm{F}$& Eq.~(\ref{Kepler}) (Keplerian) or Eq.~(\ref{BZ}) (rigid)\\
			Radius where $n$ peaks on $z=\pm z_1$& $R_p$&varied in $[0, 100r_g]$\\
			Viewing angle & $\Theta$ &$25^\circ$\\
			Width of the Gaussian ring&$\Delta$ &$5r_g$\\
			Jet shape& $\nu$ &$1$ (paraboloidal jet)\\
			Energy spectral index of the non-thermal electrons& $p$ & $1.1$\\
			Mass of the BH& $M_\mathrm{BH}$& $3.4\times10^9M_\odot$\\
			Strength of the magnetic field at $(R,z)=(0,\pm z_1)$& $Az_1^{2-\nu}$ &$100$ G \\
			Number density of the non-thermal electrons at $(R,z)=(R_p,\pm z_1)$ & $n_0$&$1$ cm${}^{-3}$ \\
			Dimensionless Kerr parameter of the BH&$a$&varied in $[0, 0.998]$\\
			Height of the plane where $n$ is given in a ring shape by Eq.~(\ref{ntext}) & $z_1$ & $5r_g$\\
			Minimal Lorentz factor of the non-thermal electrons & $\gamma _\mathrm{min}'$& $100$ \\
			Maximal Lorentz factor of the non-thermal electrons & $\gamma _\mathrm{max}'$& $\infty$ \\
			Observational frequency& $\omega/(2\pi)$ & $44$ GHz\\
			Luminosity distance to the jet & $D$ & 16.7 Mpc\\
			Inclination of the projected jet axis measured from the east direction& & $20^\circ$ toward north-east\\
			Beam kernel & & Walker et al. (2008) (VLBA)
			\\ \hline
		\end{tabular}
	\end{center}
\end{table*}

\subsection{Model Parameters}\label{method.para}
In our force-free model, there remain 10 parameters: $\Omega _\mathrm{F}$, $R_p$, $\Theta$, $\Delta$, $\nu$, $p$, $M_\mathrm{BH}$, $A$, $n_0$, and $\omega$. We systematically change them and investigate the effects on our synthetic radio images. The former two parameters ($\Omega _\mathrm{F}$ and $R_p$) are especially important, since they can drastically change radio images as shown in Sec.~\ref{sec.results}. The viewing angle $\Theta$ is found to be less important for limb-brightened features while it can be important for the brightness ratio between the jet and counter jet (See Sec.~\ref{sec.discussion.view}). The choices of the other parameters do not qualitatively alter the synthetic images (See Appendix~\ref{sec.dependence}).

We consider two patterns of $\Omega _\mathrm{F}$.
The first choice of $\Omega _\mathrm{F}$ is the same as in \citetalias{BL}, where the magnetic field is threaded through a razer-thin accretion disk at the equatorial plane. Since the field rotates with the disk, $\Omega _\mathrm{F}$ is given by 
\begin{equation}
\label{Kepler}
\Omega_\mathrm{F} =\left\{
\begin{array}{lc}
\Omega_\mathrm{Kep}(\tilde{R})& (\tilde{R}>R_\mathrm{ISCO}) \\
\Omega_\mathrm{Kep}(R_\mathrm{ISCO})& (\tilde{R}\le R_\mathrm{ISCO})
\end{array}\right.,
\end{equation}
where $\tilde{R}$ is the foot point radius  of a given magnetic field line measured on the equatorial plane and $R_\mathrm{ISCO}$ is the radius of the innermost stable circular orbit (ISCO) for prograde rotations. $\Omega_\mathrm{Kep}$ is the Keplerian angular frequency given by the dimensionless Kerr parameter, $a$, as follows \citep{Bardeen72}:
\begin{equation}
\label{KeplerFomula}
\Omega_\mathrm{Kep} = \frac{\sqrt{GM_\mathrm{BH}}}{\sqrt{R^3}+a\sqrt{r_G^3}},
\end{equation}
where $r_G := r_g/2 = GM_\mathrm{BH}/c^2$ is the gravitational radius.

The other choice of $\Omega _\mathrm{F}$ is motivated by the BZ process, which was not considered in \citetalias{BL}. In the BZ process, $\Omega_\mathrm{F}$ is nearly a constant given by 
\begin{equation}\label{BZ}
\Omega_\mathrm{F} = \frac{1}{2}\Omega_\mathrm{BH} = \frac{ac}{4r_+},
\end{equation}
where $\Omega_\mathrm{BH}$ is the rotational frequency of the Kerr BH and $r_+=(1+\sqrt{1-a^2})r_G$ is the horizon radius \citep{BZ,MG04}. 
We assume that the shape of magnetic field lines changes above the equator so that the field lines penetrate the event horizon while the shape far away from the equator is given by Eq.~(\ref{eq.Psi}). Such a field configuration may be possible, depending on the profile of the external pressure of the disk wind and/or corona, which collimates the jet and is responsible for the global jet shape \citep{McKinney06,Nakamura13}. Since the jet structure near the central region is not important for the limb-brightened feature observed far from the central BH, we use Eq.~(\ref{eq.Psi}) in the entire region for simplicity.

The ring radius ($R_p$) is systematically changed from $0$ to some sufficiently large value. For comparison to \citetalias{BL}, the other parameters are fixed to the fiducial values in \citetalias{BL}: $M_\mathrm{BH}=3.4 \times 10^9 M_\odot$, $\Delta = 5r_g$, $\nu = 1$, $p=1.1$, and $\Theta=25^\circ$, which were chosen for the M87 jet. Accordingly, we henceforth consider M87, which is an example of AGNs that show a symmetrically limb-brightened jet with a dim counter jet.\footnote{As a first step, we investigate the relations between the jet images and the important jet parameters ($\Omega_\mathrm{F}$ and $R_p$) while fixing other parameters to the fiducial values and try to produce radio images with a symmetrically limb-brightened jet and a dim counter jet. We do not try to find the best-fit parameters for the M87 jet images.} We calculate jet images in $\sim$ several~mas scale, where the limb-brightened feature of the jet is observed with VLBI \citep{Ly07,Walker08,Hada11,Hada13,Hada16,Mertens16}. The fiducial mass of the BH leads to 1~mas $\sim$ $0.08$~pc $\sim250r_g$ for $D=16.7$~Mpc \citep{D}. We use the beam kernel for VLBA given in \citet{Walker08} and assume that the M87 jet is inclined toward north-east by $20^\circ$ measured from the east direction on the sky. We note that $A$ and $n_0$ are related only to the normalization of the intensity. We adopt the following values throughout the paper: $Az_1^{2-\nu}=100$~G, which corresponds to the strength of the magnetic field at $(R,z)=(0,\pm z_1)$ \citep{Kino15}, and $n_0=1$~cm${}^{-3}$, which produces a peak intensity that is roughly consistent with observations of M87 in order of magnitude for our best model described in Sec.~\ref{sec.FastBZ}.\footnote{As seen in the right panel in Fig.~\ref{fig.BZFast}, the peak intensity $\sim 10^3$~milli-Jansky per beam for $R_p = 40r_g$ is roughly consistent with those observed in M87 \citep[$\sim 5\times 10^2$~milli-Jansky per beam,][]{Hada16} in order of magnitude. The peak intensity would reduced for more realistic model, since our model assumes optically thin jets while the central core of the M87 jet is actually optically thick for 44 GHz.} A typical value of the magnetization factor $\sigma=B^2/(4\pi\Gamma nm_pc^2)$, where $m_p$ is the proton mass, is given by
\begin{equation}\label{eq.sigma}
	\sigma \sim 5.3 \times 10^5 \Gamma ^{-1}\left(\frac{B}{100~\mathrm{G}}\right)^2\left(\frac{n}{1~\mathrm{cm}^{-3}}\right)^{-1}.
\end{equation}
In fact, the force-free approximation $\sigma \gg 1$ consistently holds for our jet models as shown in the next section. We use the observed frequency of $\omega/(2\pi)=44$~GHz to synthesize the intensity maps while the choice of $\omega$ does not affect the shape of radio contour maps under the optically thin assumption.\footnote{The optically-thin assumption holds well for this frequency for the jet models in the main body of this paper, which correspond to Figs.~\ref{fig.Kepler}, \ref{fig.BZFast}, \ref{fig.BZSlow}, and \ref{fig.Theta}.} We summarize the model parameters in Table~1. In Appendix~\ref{sec.dependence}, some of the above parameters are varied around our fiducial values to study the effects on radio images whereas they do not change our conclusions.

\section{Results}\label{sec.results}

\begin{figure*}
	\centering \includegraphics[bb = 0 0 1620 1166, width=\textwidth]{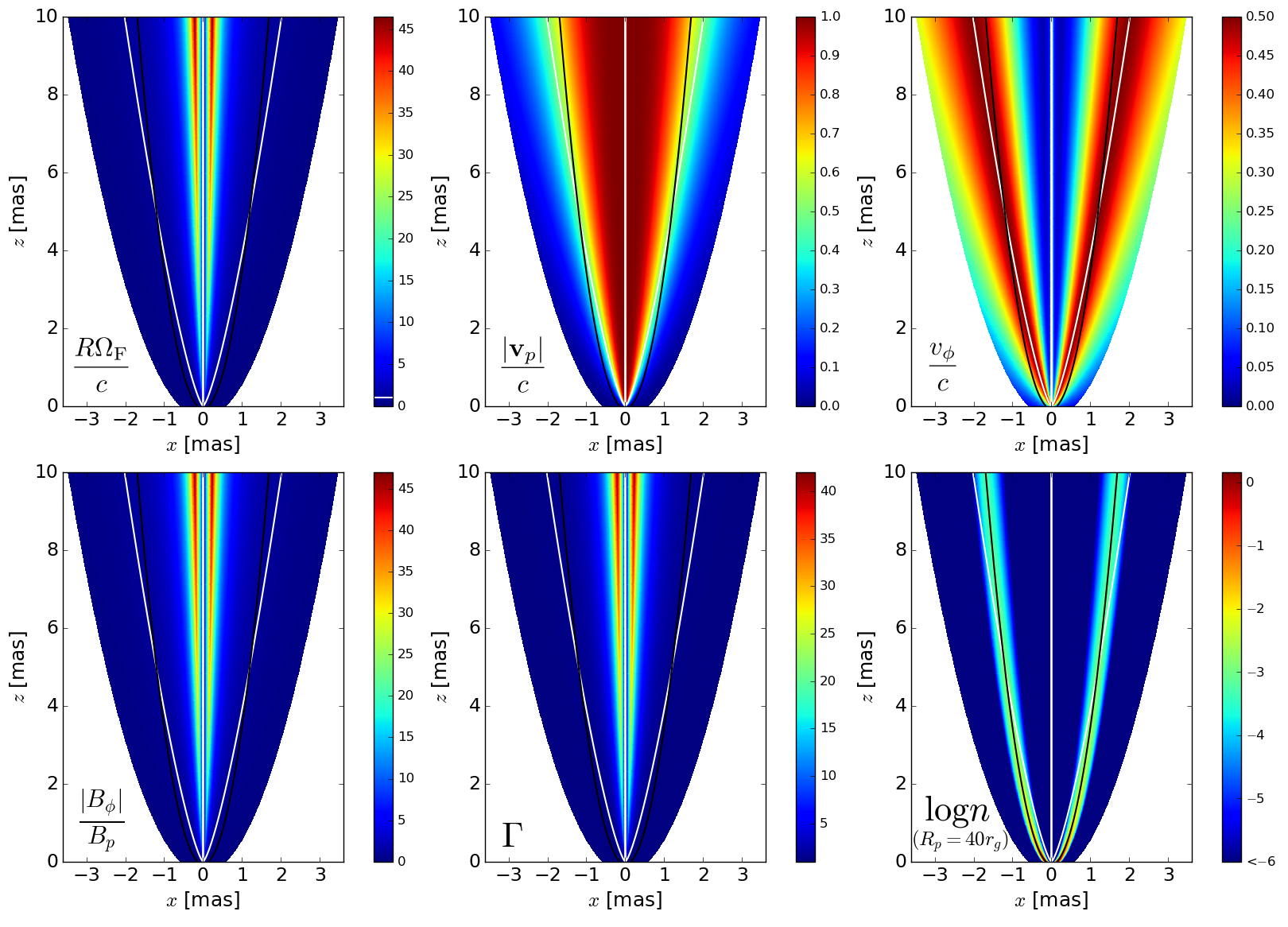}
	\caption{Physical quantities in the jet $(y = 0$, $z>0)$ for the Case 1. The jet is axisymmetric around the $z$-axis and the BH exists at the origin. The jet structure is symmetric with respect to the equatorial plane ($z=0$). Upper-left: the color map of $R\Omega_\mathrm{F}/c$. Upper-middle: The poloidal speed normalized by the speed of light, $|{\bf v}_p| /c$. Upper-right: The azimuthal speed normalized by the speed of light, $v_\phi/c$. Lower-left: The ratio of the toroidal and poloidal magnetic field strengths, $|B_\phi|/B_p$. Lower-middle: The Lorentz factor, $\Gamma$. Lower-right: The number density of the non-thermal electrons for $R_p=40r_g$ in the logarithmic scale, where the region with $\log n<-6$ is filled with the same color for the visibility of dense region. In these panels, the white lines are the light cylinder and the black ones are the poloidal magnetic filed lines that pass through $R = 40 r_g$ at $z=5r_g$. The jet was cut out along the magnetic field surface that goes through $R=150r_g$ at $z = 5 r_g$. Note that 1~mas corresponds to $\sim 250r_g$ (i.e., $1r_g \sim 4\times10^{-3}$~mas).}
	\label{fig.fieldKepler}
\end{figure*}

\begin{figure*}
	\begin{tabular}{cc}
		\begin{minipage}{0.45\hsize}
			\begin{center}
				\includegraphics[bb = 0 0 496 420, width=\textwidth]{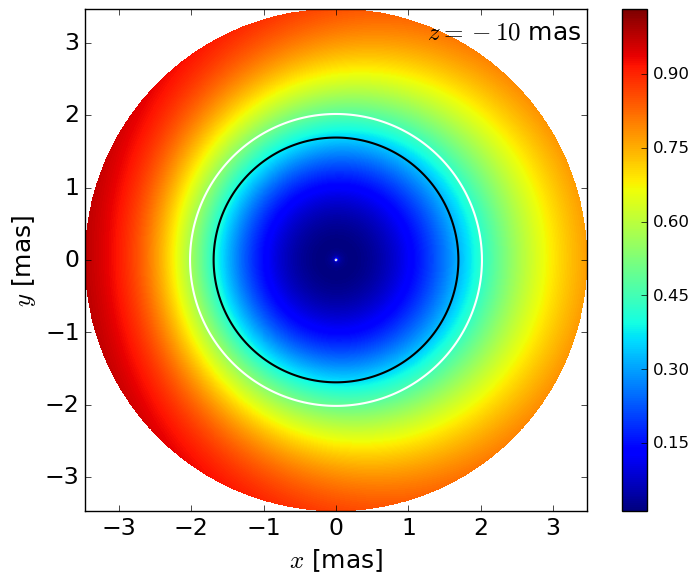}
			\end{center}
		\end{minipage} &
		\begin{minipage}{0.45\hsize}
			\begin{center}
				\includegraphics[bb = 0 0 492 419, width=\textwidth]{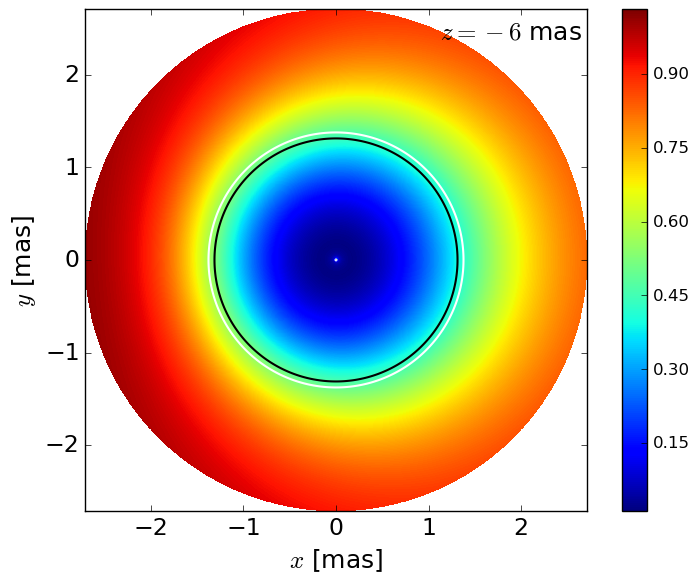}
			\end{center}
		\end{minipage} \\
				\begin{minipage}{0.45\hsize}
			\begin{center}
				\includegraphics[bb = 0 0 519 420, width=\textwidth]{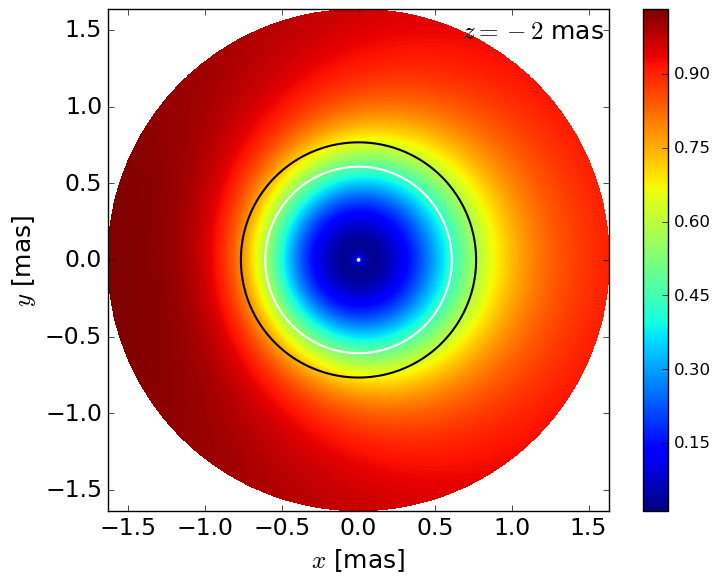}
			\end{center}
		\end{minipage} &
		\begin{minipage}{0.45\hsize}
			\begin{center}
				\includegraphics[bb = 0 0 511 420, width=\textwidth]{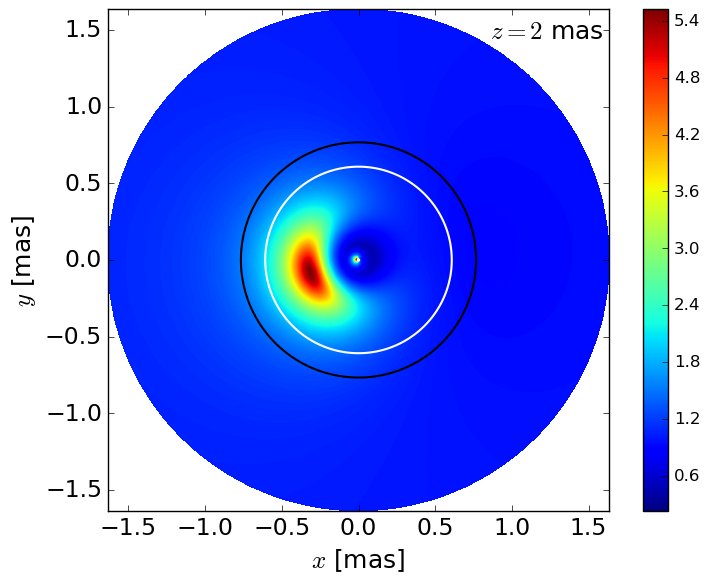}
			\end{center}
		\end{minipage} \\
		\begin{minipage}{0.45\hsize}
			\begin{center}
				\includegraphics[bb = 0 0 496 420, width=\textwidth]{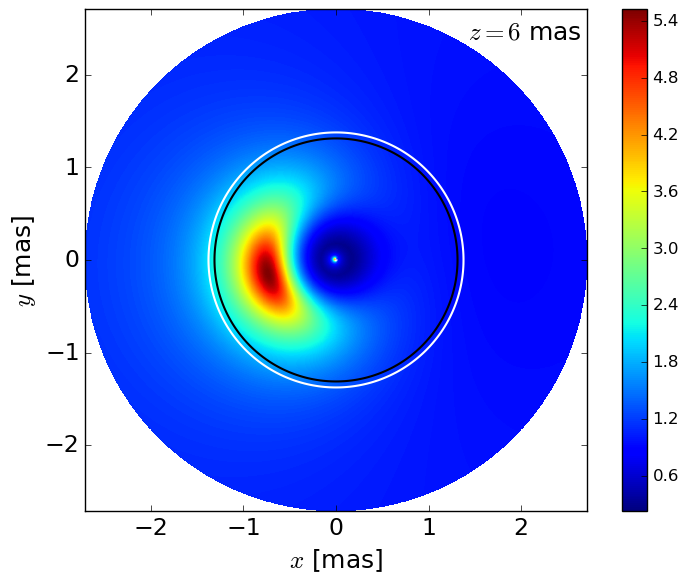}
			\end{center}
		\end{minipage} &
		\begin{minipage}{0.45\hsize}
			\begin{center}
				\includegraphics[bb = 0 0 496 420, width=\textwidth]{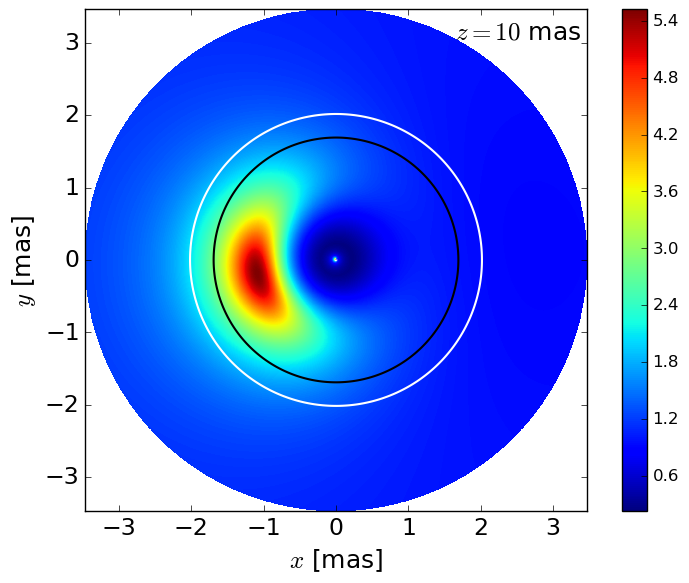}
			\end{center}
		\end{minipage} 
	\end{tabular}
	\caption{The beaming factor $\delta = 1/[\Gamma(1-\beta\mu)]$ for the observer with $\Theta=25^\circ$ on some horizontal slices of the approaching $(z>0)$ and counter $(z<0)$ jets of the Case~1. The sliced plane is designated in the upper-right corner in each panel. As same as in Fig.~\ref{fig.fieldKepler}, the white and black lines indicate the light cylinder and the magnetic filed surface that passes through $R = 40 r_g$ at $z=5r_g$. Note that 1~mas corresponds to $\sim 250r_g$ (i.e., $1r_g \sim 4\times10^{-3}$~mas).}
	\label{fig.beam_Kepler}
\end{figure*}

\begin{figure*}[!t]
	\centering \includegraphics[bb = 0 0 708 418, width=\textwidth
	]{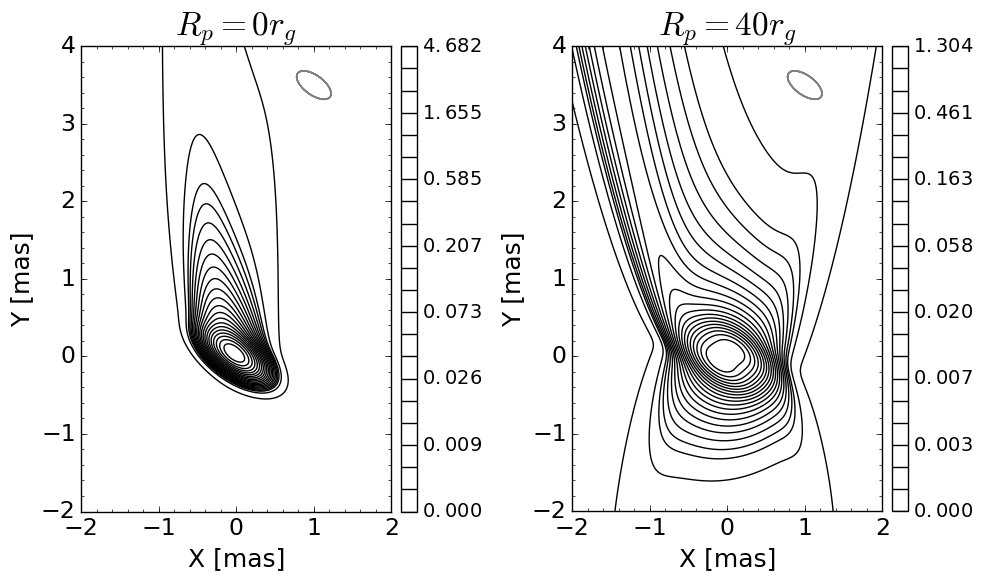}
	\caption{Radio intensity maps for the Case 1, where the magnetic field penetrates the Keplerian accretion disk. The unit of the intensity is milli-Jansky per beam. The contours are drawn as follows: The inner $20$ contours are for $\sqrt{2}^{-k}$ ($k=0,\cdots,19$) while the outermost two are for $\sqrt{2}^{-21}$ and $0.1\sqrt{2}^{-21}$, respectively. The $Y$-axis coincides with the projected jet axis and the origin is the projected location of the BH. The particle distributions are given by $R_p=0$ and $40r_g$ in the left and right, respectively, as designated above each panel. The beam shape is also plotted in gray at the top right corner in each panel. Note that 1~mas corresponds to $\sim 250r_g$ (i.e., $1r_g \sim 4\times10^{-3}$~mas).}
	\label{fig.Kepler}
\end{figure*}

\subsection{Case 1: Disk-threaded Model}
First, we examine the case of the disk-threaded model, in which the magnetic fields penetrate the Keplerian accretion disk and $\Omega_\mathrm{F}$ is given by Eq.~(\ref{Kepler}). We here show the results for $a=0.998$, which is the fiducial value in \citetalias{BL}, since those for smaller $a$ are qualitatively the same. 
This model has $R_\mathrm{ISCO}\sim1.2r_+\sim0.62r_g$ and $\Omega_\mathrm{Kep}(R_\mathrm{ISCO})\sim2.5\times10^{-5}$~s$^{-1}$.

The jet structure of this model is presented in Fig.~\ref{fig.fieldKepler}.\footnote{The intrinsic jet length of $10$~mas corresponds to the projected length of $\sim 4.2$~mas for the viewing angle of $\Theta=25^\circ$.} The upper-left panel shows the distribution of an important quantity, $R\Omega_\mathrm{F}/c$ (See also Fig.~\ref{fig.closeup} for the close-up around the origin). As shown by the white lines, the so-called light `cylinder', where $R\Omega_\mathrm{F}/c=1$ is satisfied, forms not only a vertical surface around the jet axis but also a curved one far from the jet axis \citep{Blandford76}. The former truncated cylinder is formed at $R=R_\mathrm{lc,1}:=c/\Omega_\mathrm{Kep}(R_\mathrm{ISCO}) \sim1.2r_g\sim4.8\times10^{-3}$~mas due to the uniform rotation of the magnetic field passing through inside the ISCO. The latter curved surface ($z\propto R^{4/3}$ at $R\gg r_G$; See Appendix~\ref{sec.LC}) is, on the other hand, attributed to the differential rotation of the magnetic field lines anchored to the accretion disk. As a result, $R\Omega_\mathrm{F}/c$ exceeds unity only in a limited region bound by these two surfaces. This means that the jet edge part is dominated by poloidal magnetic field and not efficiently accelerated to poloidal directions as shown in the lower-left and upper-middle panels in Fig.~\ref{fig.fieldKepler} (See Eqs.~(\ref{accele1}) and (\ref{non-accele1}) for asymptotic relations between $|{\bf v}_p|$ and $R\Omega_\mathrm{F}/c$). The highly relativistic poloidal speed is realized, on the other hand, only at $R\gtrsim R_\mathrm{lc,1}$, which is near the jet axis. The jet rotational speed, $v_\phi$, is shown in the upper-right panel in Fig.~\ref{fig.fieldKepler}. $v_\phi$ peaks $\sim 0.5c$ around the light cylinder and reduces apart from it as indicated in Eqs.~(\ref{accele2}) and (\ref{non-accele2}). The lower-middle panel in Fig.~\ref{fig.fieldKepler} shows the Lorentz factor, which manifestly shows the jet is relativistic only near the jet axis as explained above. The lower-right panel shows the number density of the non-thermal electrons for $R_p=40r_g$ (one of the fiducial cases) in the logarithmic scale. As is evidently, the non-thermal electrons are concentrated on the magnetic field lines that pass through $R=40r_g$ at $z=5r_g$ ($\Psi/A\sim35.3r_g$; drawn by black lines in Fig.~\ref{fig.fieldKepler}) while the number density rapidly decreases away from the lines. $n$ is also reduced along a field line upward. We note that the magnetization factor $\sigma$ given by Eq.~(\ref{eq.sigma}) is low at the dense region around the black lines but is much larger than unity in the displayed region ($\sigma \gtrsim 8\times10^3$), which ensures the use of the force-free approximation for this model.

Figure~\ref{fig.beam_Kepler} shows the beaming factor $\delta := 1/[\Gamma (1-\beta \mu)]$ on some horizontal slices of the jet. The top and middle-left panels are for the counter jet ($z<0$) and the others are for the jet ($z>0$) that directs to the observer with the viewing angle of $\Theta=25^\circ$. The beaming effect becomes remarkable in the region with $R\Omega_\mathrm{F}/c \gg 1$, where the poloidal speed becomes relativistic and the Lorentz factor is large. In the approaching jet side, the distribution of $\delta$ is highly asymmetric due to the jet rotation, which reaches $\sim 0.5c$ around the curved light `cylinder' as presented in the upper-right panel in Fig.~\ref{fig.fieldKepler}. Through $\mu$ in Eq.~(\ref{jtext}), jet rotations lead to the opposite effects of relativistic beaming in the left and right sides of the jet. The left side of the jet is coming to the observer and, as a result, strongly beams light toward the observer whereas the right side of the jet is going away from the observer and, hence, does not efficiently beam light to the observer. We note that the peaks of $\delta$ and $\Gamma$ in each slice do not necessarily coincide due to the misalignment of the observer and flow directions. In the counter-jet side, on the other hand, $\delta$ is suppressed below unity almost in the entire region. The suppression is especially strong in the region with $R\Omega_\mathrm{F}/c \gg 1$ and the asymmetry of $\delta$ due to the jet rotation is also seen as in the approaching jet side.

The left panel in Fig.~\ref{fig.Kepler} shows the calculated radio image for $R_p = 0$. This model is essentially the same as the standard (M0) model in \citetalias{BL}, where the non-thermal electrons cling to the jet axis. As expected, the jet axis is the brightest due to the concentration of electrons and any limb-brightened feature is not seen. The counter jet is not seen in the radio map due to the relativistic beaming to the opposite direction of the observer. We also note that the radio intensity is larger in the left hand side of the jet in the figure because of the asymmetric beaming effect shown in Fig.~\ref{fig.beam_Kepler}.

We here pick up the result for $R_p=40r_g$, while the results for $R_p>0$ are qualitatively the same as mentioned later. In this example, the non-thermal electrons are nearly on the curved light-cylinder surface for $|z|\le 10$~mas as indicated by the magnetic field lines with $\Psi/A\sim35.3r_g$ (the black lines in Figs.~\ref{fig.fieldKepler} and \ref{fig.beam_Kepler}), which reasonably trace the dense region of the non-thermal electrons. The right panel in Fig.~\ref{fig.Kepler} shows the synthesized radio image for $R_p=40r_g$. One of the most striking features is the strongly asymmetric limb brightening: The left hand side of the jet axis in the figure (i.e., the northern part on the sky) is more luminous than the counterpart in the right hand side (i.e., the southern part). The limb brightening is understood just as a reflection of the assumed $R_p$. The large asymmetry is, on the other hand, due to the rotation of the jet.
As seen in Fig.~\ref{fig.beam_Kepler}, the asymmetry of the beaming factor in the approaching jet side is relatively large near the magnetic field line $\Psi/A\sim35.3r_g$, where $v_\phi$ reaches ($\sim 0.5c$) as plotted in the upper-right panel in Fig.~\ref{fig.fieldKepler}. We note that the synchrotron emission is intrinsically asymmetric even in the fluid rest frame, since the pitch angle of the relativistic electrons that direct to the observer is different between the right and left sides of the jet due to winding magnetic field lines, which is included in Eq.~(\ref{jd}) through $\sin \psi' ({\bf n'})$. The intrinsic asymmetry is, however, found to be minor compared to the asymmetry induced by the relativistic beaming.

The luminous counter jet is another notable feature for $R_p=40r_g$ as seen in the right panel in Fig.~\ref{fig.Kepler}. The counter jet becomes apparent, in contrast to the observations of the M87 jet \citep[e.g.][]{Hada16}, since the relativistic boost to poloidal directions is so weak. As shown in the upper-middle panel in Fig.~\ref{fig.fieldKepler}, the poloidal speed on the magnetic field line $\Psi/A\sim 35.3r_g$ is relatively small $\lesssim 0.7c$. As a result of the decrease of $|{\bf v}_p|$ toward the jet edge, $\delta$ increases to unity toward the jet edge in the counter-jet side as shown in Fig.~\ref{fig.beam_Kepler}. $\delta$ is $\sim 0.5$ on the magnetic field line $\Psi/A\sim 35.3r_g$, which is not sufficient to darken the counter jet.

The results for other $R_p>0$ are qualitatively the same: We confirmed that the radio images still keep the strong asymmetry for $0 < R_p < 40r_g$, as indicated by the asymmetric candle-flame-like image with the brighter northern edge for $R_p=0$ in the left panel in Fig.~\ref{fig.Kepler}. The counter jet for $0 < R_p < 40r_g$ becomes less luminous than for $R_p=40r_g$ owing to larger $|{\bf v}_p|$. We also found for $R_p>40r_g$ that the asymmetry of the limb brightening can be weaker thanks to smaller asymmetry of $\delta$ between the right and left sides of the jet, which is due to smaller $v_\phi$ but the counter jet becomes more luminous due to smaller $|{\bf v}_p|$.

\begin{figure*}
	\centering \includegraphics[bb = 0 0 1620 1166, width=\textwidth]{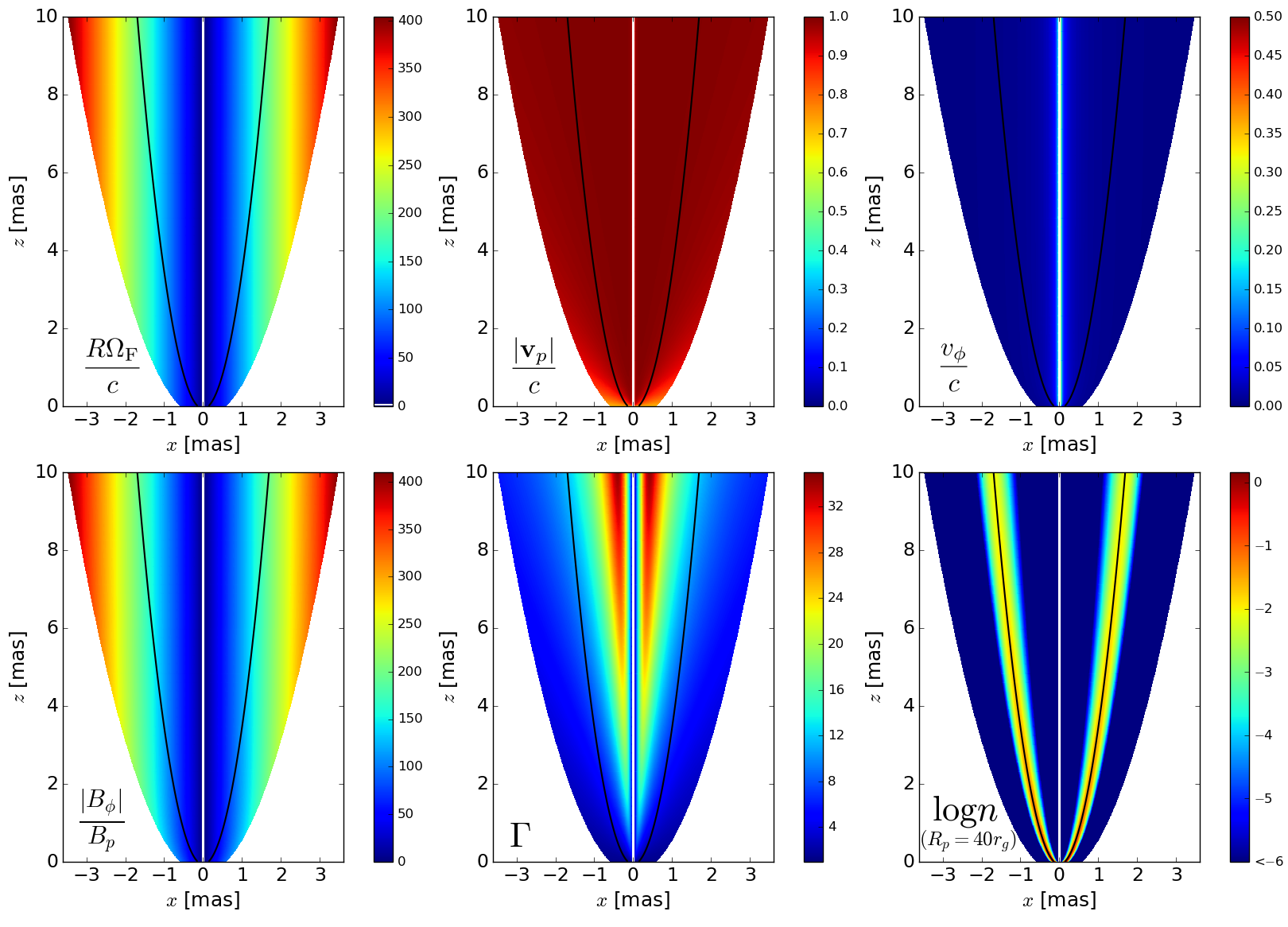}
	\caption{Same as Fig.~\ref{fig.fieldKepler} but for the Case 2 with $a=0.998$.}
	\label{fig.fieldFastBZ}
\end{figure*}

\begin{figure*}
	\begin{tabular}{cc}
		\begin{minipage}{0.45\hsize}
			\begin{center}
				\includegraphics[bb = 0 0 492 419, width=\textwidth]{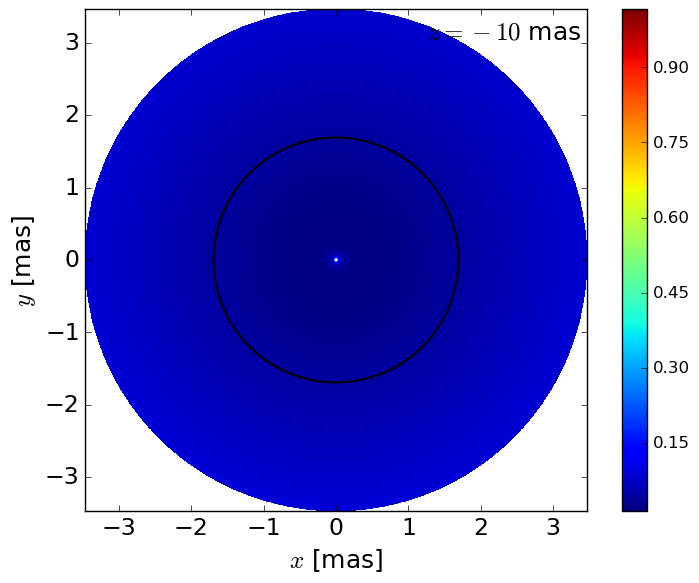}
			\end{center}
		\end{minipage} &
		\begin{minipage}{0.45\hsize}
			\begin{center}
				\includegraphics[bb = 0 0 492 419, width=\textwidth]{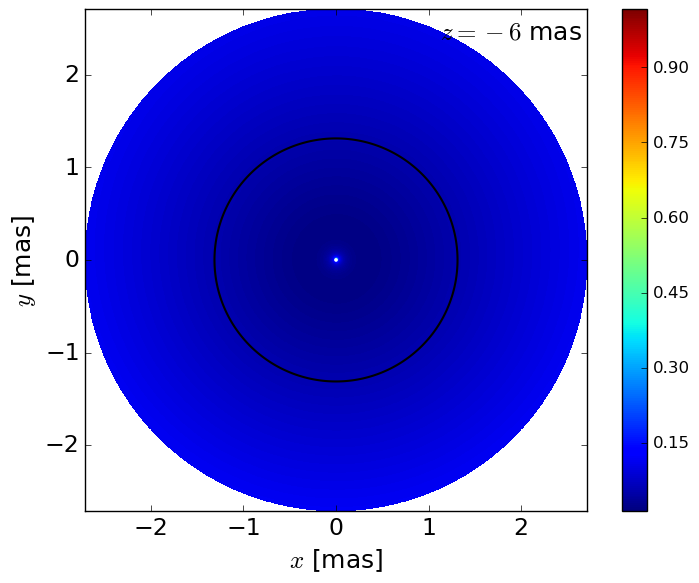}
			\end{center}
		\end{minipage} \\
		\begin{minipage}{0.45\hsize}
			\begin{center}
				\includegraphics[bb = 0 0 507 419, width=\textwidth]{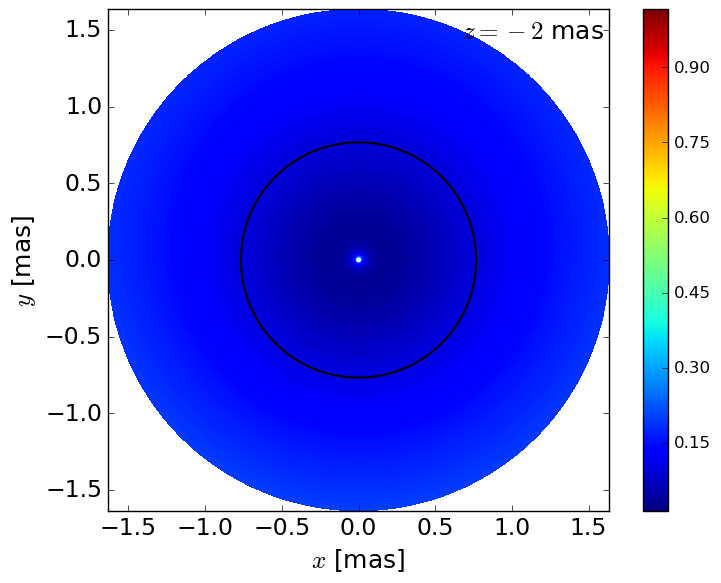}
			\end{center}
		\end{minipage} &
		\begin{minipage}{0.45\hsize}
			\begin{center}
				\includegraphics[bb = 0 0 511 420, width=\textwidth]{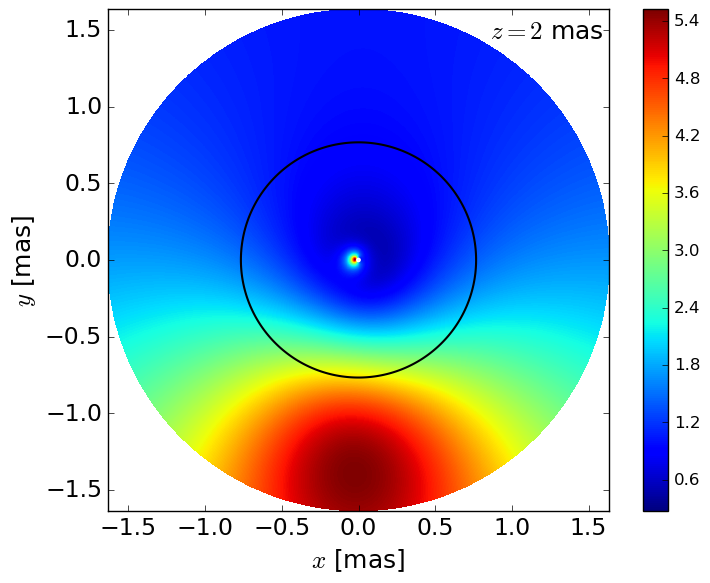}
			\end{center}
		\end{minipage} \\
		\begin{minipage}{0.45\hsize}
			\begin{center}
				\includegraphics[bb = 0 0 496 420, width=\textwidth]{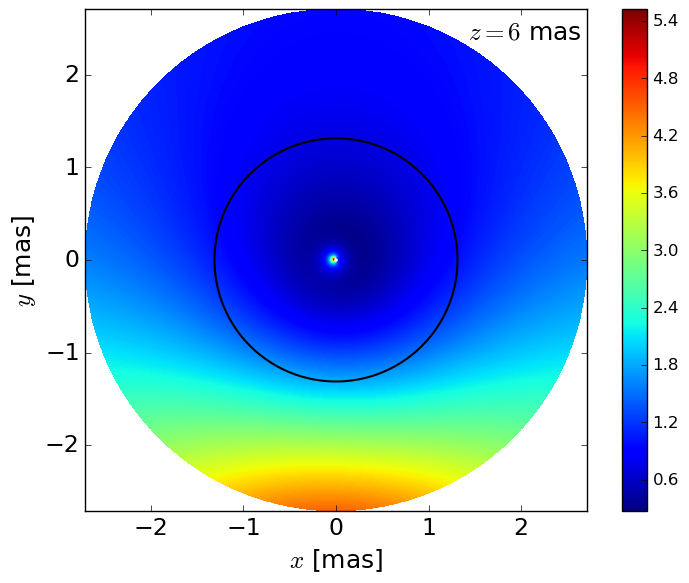}
			\end{center}
		\end{minipage} &
		\begin{minipage}{0.45\hsize}
			\begin{center}
				\includegraphics[bb = 0 0 495 419, width=\textwidth]{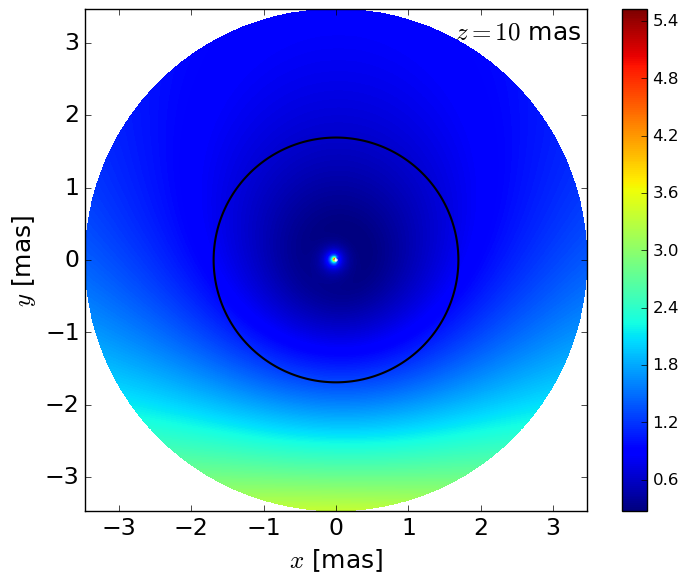}
			\end{center}
		\end{minipage} 
	\end{tabular}
	\caption{Same as Fig.~\ref{fig.beam_Kepler} but for the Case 2 with $a=0.998$.}
	\label{fig.beam_FastBZ}
\end{figure*}

\begin{figure*}
	\centering \includegraphics[bb = 0 0 708 418, width=\textwidth
	]{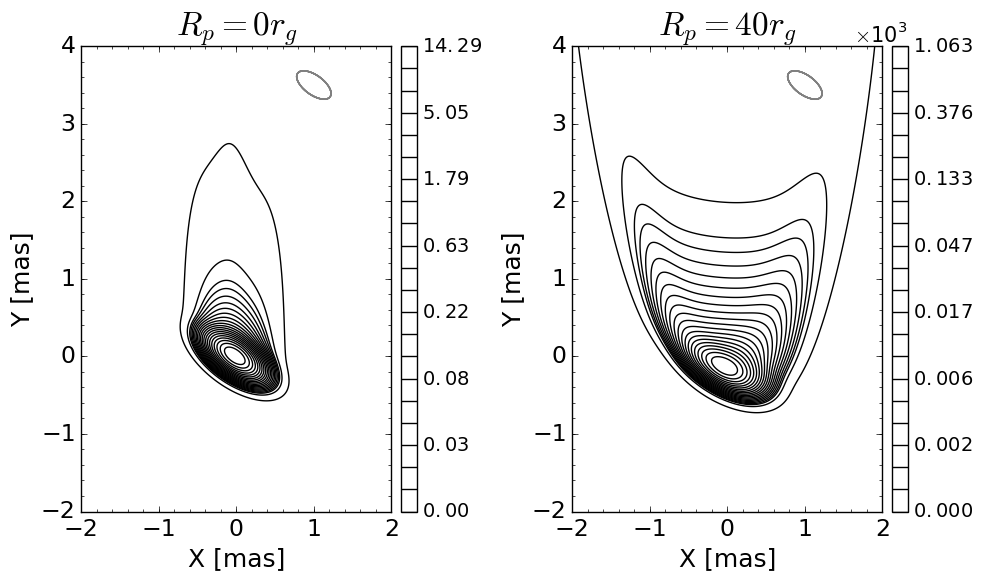}
	\caption{Same as Fig.~\ref{fig.Kepler} but for the Case 2 with $a=0.998$.}
	\label{fig.BZFast}
\end{figure*}

\begin{figure*}
	\centering \includegraphics[bb = 0 0 1620 1166, width=\textwidth]{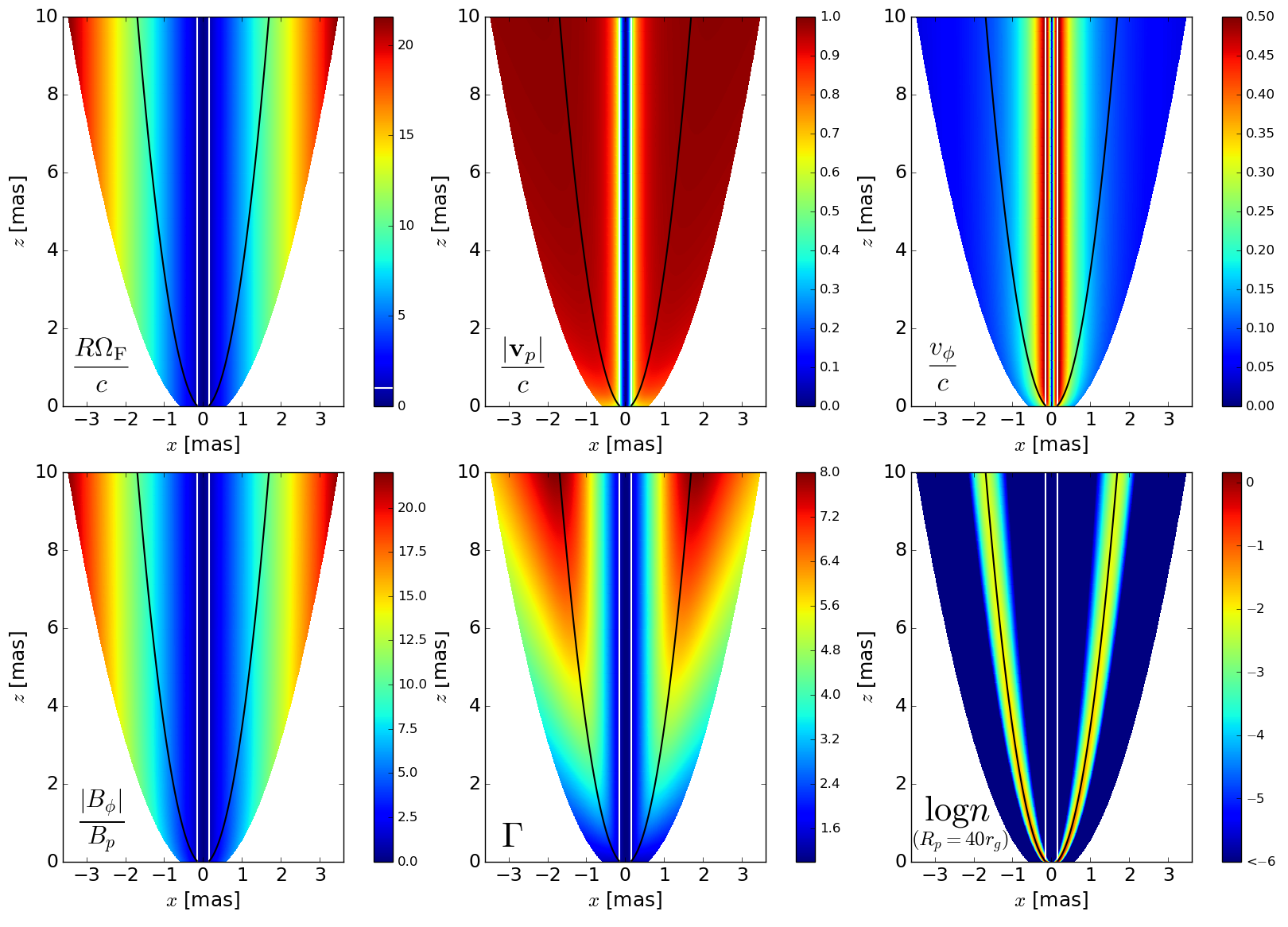}
	\caption{Same as Fig.~\ref{fig.fieldKepler} but for the Case 2 with $a=0.1$.}
	\label{fig.fieldSlowBZ}
\end{figure*}

\begin{figure*}
	\begin{tabular}{cc}
		\begin{minipage}{0.45\hsize}
			\begin{center}
				\includegraphics[bb = 0 0 496 420, width=\textwidth]{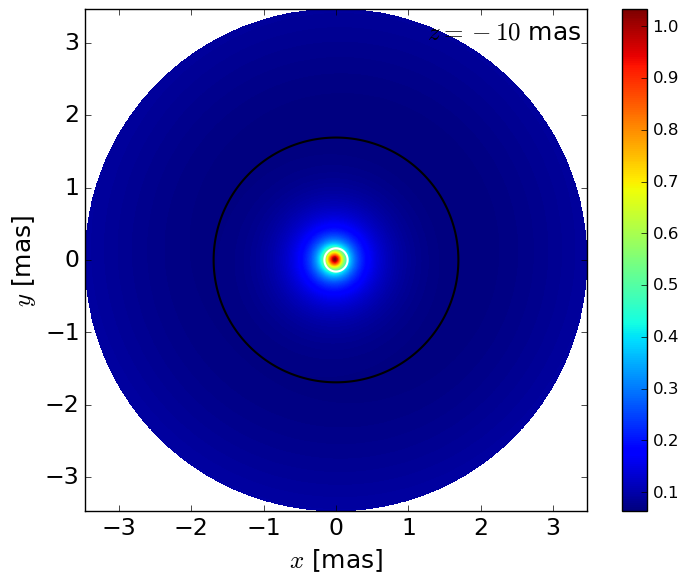}
			\end{center}
		\end{minipage} &
		\begin{minipage}{0.45\hsize}
			\begin{center}
				\includegraphics[bb = 0 0 496 420, width=\textwidth]{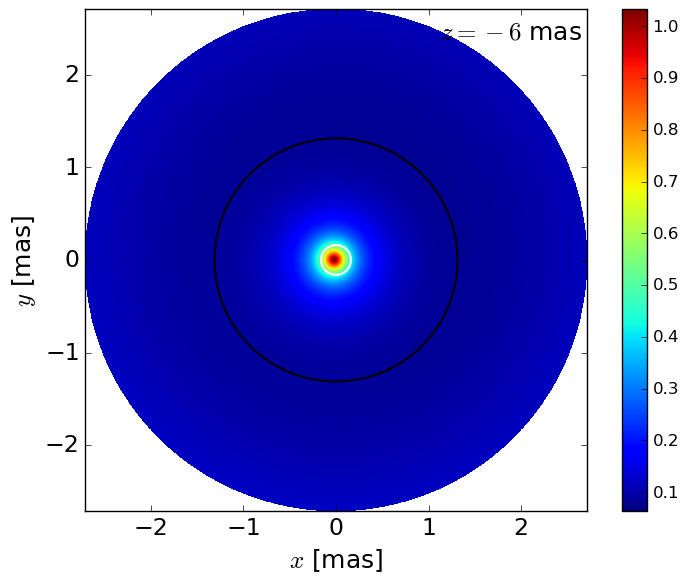}
			\end{center}
		\end{minipage} \\
		\begin{minipage}{0.45\hsize}
			\begin{center}
				\includegraphics[bb = 0 0 511 420, width=\textwidth]{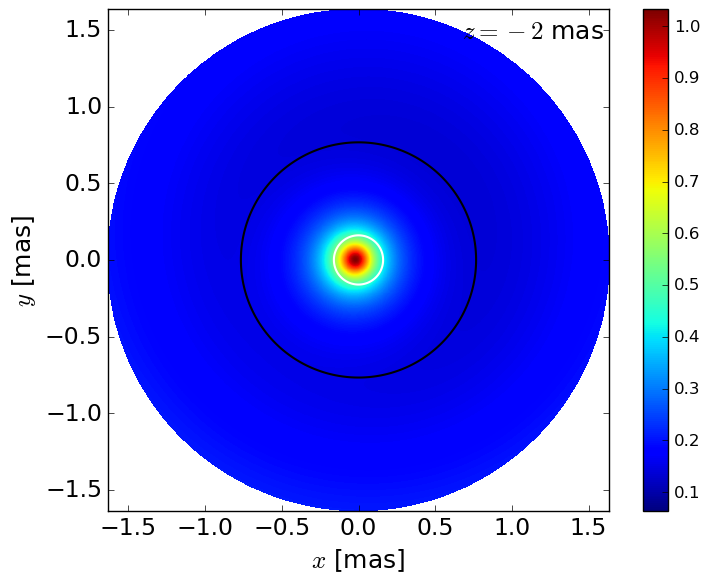}
			\end{center}
		\end{minipage} &
		\begin{minipage}{0.45\hsize}
			\begin{center}
				\includegraphics[bb = 0 0 511 420, width=\textwidth]{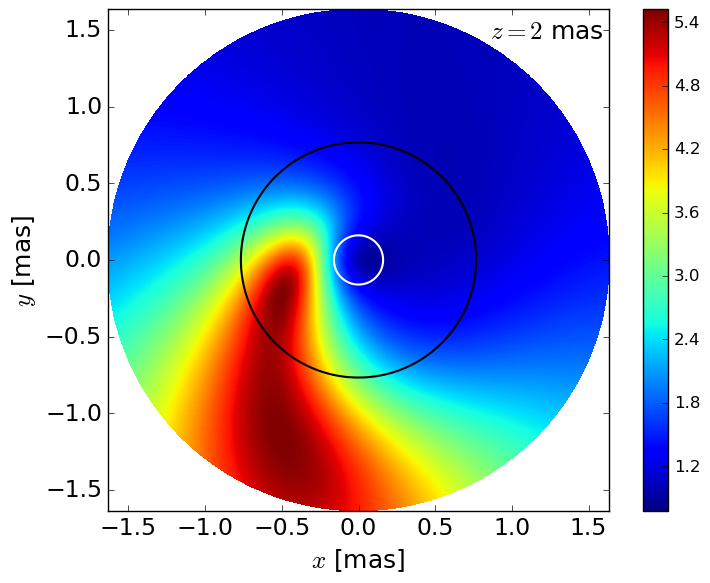}
			\end{center}
		\end{minipage} \\
		\begin{minipage}{0.45\hsize}
			\begin{center}
				\includegraphics[bb = 0 0 496 420, width=\textwidth]{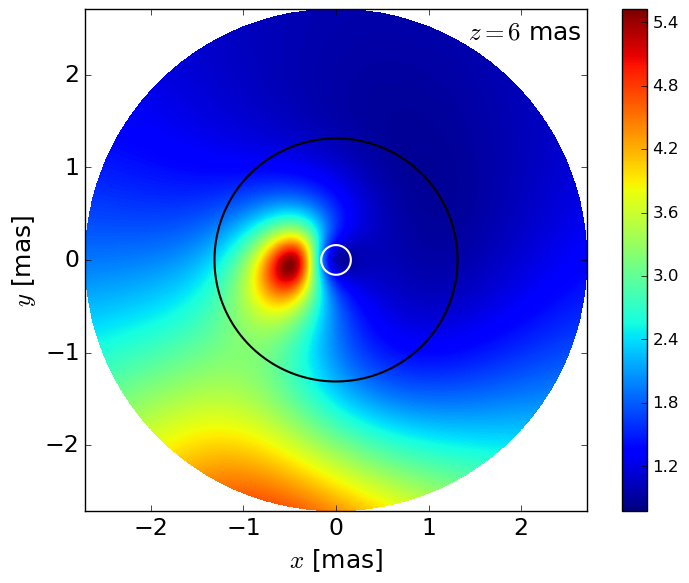}
			\end{center}
		\end{minipage} &
		\begin{minipage}{0.45\hsize}
			\begin{center}
				\includegraphics[bb = 0 0 496 420, width=\textwidth]{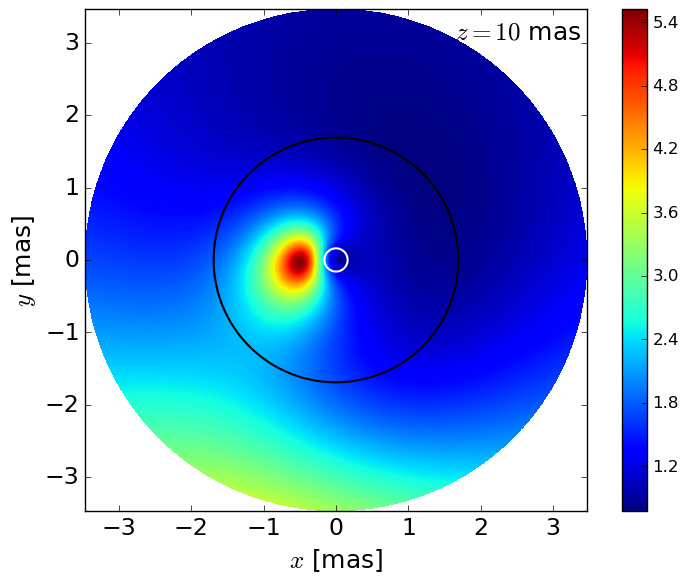}
			\end{center}
		\end{minipage} 
	\end{tabular}
	\caption{Same as Fig.~\ref{fig.beam_Kepler} but for the Case 2 with $a=0.1$.}
	\label{fig.beam_SlowBZ}
\end{figure*}

\begin{figure*}[!t]
	\centering \includegraphics[bb = 0 0 708 412, width=\textwidth
	]{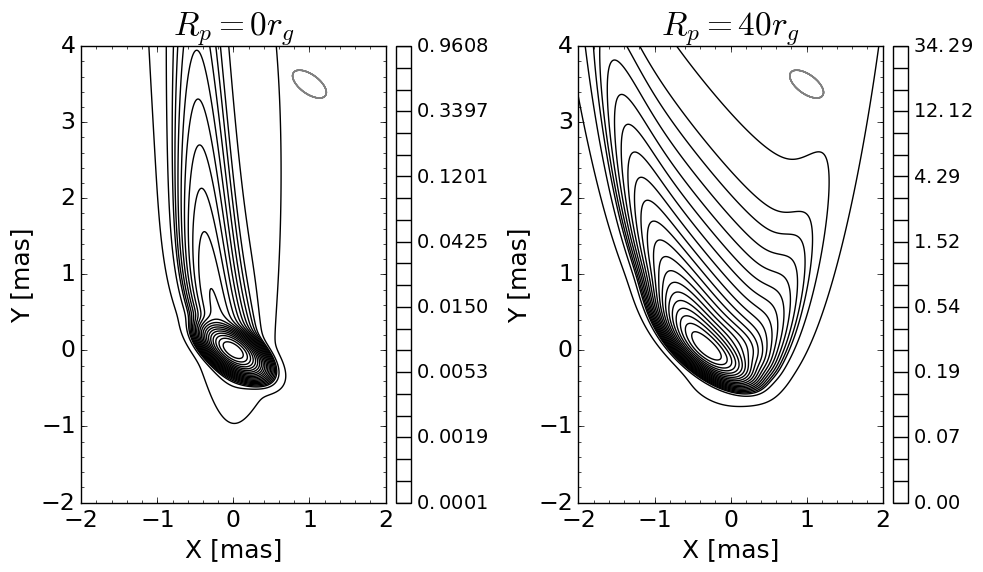}
	\caption{Same as Fig.~\ref{fig.Kepler} but for the Case 2 with $a=0.1$.}
	\label{fig.BZSlow}
\end{figure*}

\subsection{Case 2: BH-threaded Model}
We investigate the other case, where $\Omega_\mathrm{F}$ is a constant given by Eq.~(\ref{BZ}), motivated by the magnetic field lines penetrating the BH. The Kerr parameter is crucial in this case, since it directly controls $\Omega_\mathrm{F}$. We thus systematically study the dependence of the radio image on the Kerr parameter as well as the effect of $R_p$. We pick up two extreme cases of $a=0.998$ and $a=0.1$ as best examples.

\subsubsection{Fast-spinning BH}\label{sec.FastBZ}
First, we show the results for $a = 0.998$, for which $\Omega_\mathrm{F}$ is $1.4\times10^{-5}$~s$^{-1}$. The upper-left panel in Fig.~\ref{fig.fieldFastBZ} shows the color map of $R\Omega_\mathrm{F}/c$ in the jet. Since the magnetic field lines rigidly rotate, $R\Omega_{F}/c$ monotonically increases with $R$ and the light cylinder is given by $R=R_{\mathrm{lc,2f}}:=\Omega_{F}/c\sim 2.1r_g \sim 8.6\times 10^{-3}$~mas. We emphasize here that the jet structure is qualitatively different from those in the disk-threaded model, in which another curved surface of the light cylinder exists. The magnetic field is thus toroidally dominated in the almost entire region in the jet except for the inside of the thin light cylinder as depicted in the lower-left panel in Fig.~\ref{fig.fieldFastBZ}, which is a sharp contrast to the previous case.
As a result, the velocity field is also qualitatively different away from the jet axis: $|{\bf v}_p|$ becomes highly relativistic $(\sim c)$ and $v_\phi$ is suppressed to non-relativistic speed $(\lesssim 0.1c)$ as presented in the upper-middle and right panels in Fig.~\ref{fig.fieldFastBZ}. Around the jet axis ($R\lesssim R_\mathrm{lc,2f}$), on the other hand, the velocity field is not much different from that in the previous disk-threaded model, since the magnetic field lines near the jet axis rigidly rotate with comparable angular frequencies in these models (cf.~$\Omega_\mathrm{F}\sim 1.4\times10^{-5}$~s$^{-1}$ for this model and $\Omega_\mathrm{Kep}(R_\mathrm{ISCO})\sim2.5\times10^{-5}$~s$^{-1}$ for the previous disk-threaded model). The Lorentz factor is shown in the lower-middle panel in Fig.~\ref{fig.fieldFastBZ}. The lower-right panel exhibits $\log n$ for $R_p=40r_g$ as an example. The non-thermal electrons are concentrated on the magnetic filed lines $\Psi/A\sim35.3r_g$ (black lines), on which the magnetization factor $\sigma$ is minimized to $\sim 8\times10^5$ in the presented region but still holds a sufficiently large value for the force-free approximation. 

Figure~\ref{fig.beam_FastBZ} shows $\delta$ in the jet. As expected from the slow $v_\phi$, the difference of $\delta$ is small between the right and left sides with respect ot the observer. Due to the large $|{\bf v}_p|$, $\delta$ in the counter jet is suppressed and the radiation is strongly debeamed for the observer. In the approaching jet side, on the other hand, a part of the front side of the jet strongly beams the light to the observer ($\delta \sim 5$) whereas the back side does not due to the misalignment of the highly relativistic velocity and the observer direction. We also note that the asymmetry due to the anisotropic synchrotron radiation in the fluid rest frame is again found to be negligible.

The left panel in Fig.~\ref{fig.BZFast} displays the synthesized radio map for $R_p = 0$.
Neither a limb-brightened feature nor the counter jet is seen in the radio map as in the Case~1 with $R_p = 0$. This result is again attributed to the strong beaming effect to polar directions due to the velocity field near the jet axis.

The radio maps for $R_p>0$ can successfully show a symmetrically limb-brightened jet without a luminous counter jet as displayed in the right panel in Fig.~\ref{fig.BZFast}, where the result for $R_p=40r_g$ is shown for comparison to the counterpart in Fig.~\ref{fig.Kepler}. The symmetry of the jet image is recovered thanks to the small $v_\phi$ in the outer part of the jet away from the axis, which suppresses the beaming/debeaming asymmetry in the jet northern/southern sides as displayed in Fig.~\ref{fig.beam_FastBZ}. The counter jet is less luminous due to the highly relativistic poloidal speed, which beams the emission to the opposite direction of the observer.

For larger ring radii ($R_p > 40r_g$), the results are qualitatively the same as for $R_p=40r_g$, while the width of the jet image becomes wider. For smaller $R_p$ ($0 < R_p < 40r_g$), the jet width becomes smaller with keeping the symmetrically limb-brightened feature and gradually approaches the result for $R_p=0$.

\subsubsection{Slowly spinning BH}
We here show the results for the slowly spinning BH with $a=0.1$. The Kerr parameter results in $\Omega_\mathrm{F}=7.5\times10^{-7}$~s$^{-1}$, which is $5.3\times10^{-2}$ times as large as that for $a=0.998$ and shifts the light cylinder outward to the $\sim19$ times larger radius, $R=R_\mathrm{lc,2s}\sim 40r_g\sim0.16$~mas, as well as the other contour lines of $R\Omega_\mathrm{F}/c$ as presented in the upper-left panel in Fig.~\ref{fig.fieldSlowBZ}. The change is also reflected to the distribution of the ratio of the toroidal to poloidal magnetic field strengths as visible in the lower-left panel in Fig.~\ref{fig.fieldSlowBZ}. As a result, the region with slow poloidal speeds and fast azimuthal ones is extended from the jet axis to $R\lesssim R_\mathrm{lc,2s}$ as visible in the upper-middle and right panels in Fig.~\ref{fig.fieldSlowBZ}. In the outer part of the jet, $R\gg R_\mathrm{lc,2s}$, $v_\phi$ is increased by $\sim19$ times, compared to the case of $a=0.998$ at the same radius, since $v_\phi$ is inversely proportional to $R\Omega_{F}/c$ as given by Eq.~(\ref{accele2}). The poloidal speeds for $R\gg R_\mathrm{lc,2s}$ are not much different from those in the previous case of $a=0.998$, on the other hand, since it is asymptotically determined by the angle from the jet axis as given by Eq.~(\ref{accele1}), where $R\Omega_{F}/c$ appears in the higher order corrections. The resultant Lorentz factor is displayed in the lower-middle panel in Fig.~\ref{fig.fieldSlowBZ}, which has the asymptotically same structure as for $a=0.998$ in the jet edge part due to the dominance of the poloidal speed. The lower-right panel presents the density profile of the non-thermal electrons for $R_p=40r_g$, which are concentrated on the magnetic filed lines $\Psi/A\sim35.3r_g$ (black lines). The minimal value of $\sigma\sim 5\times10^3$ in the displayed area is consistent with the force-free assumption.

The plots of $\delta$ in Fig.~\ref{fig.beam_SlowBZ} clearly exhibits different patterns compared to the case for $a=0.998$. In the approaching jet side, $\delta$ is more asymmetric between the right and left sides due to the larger $v_\phi$. In the counter jet side, $\delta$ is still suppressed almost in the jet edge region due to highly relativistic $|{\bf v}_p|$ whereas $\delta$ is close to unity inside the light cylinder because of non-relativistic $|{\bf v}_p|$.

Figure~\ref{fig.BZSlow} shows the produced radio images for $R_p=0$ (left) and $R_p=40r_g$ (right), respectively. Most importantly, the radio image becomes highly asymmetric between the northern and southern parts due to the enhanced relativistic beaming by larger $v_\phi$, which is incompatible with the M87 jet. We also note that the counter jet becomes more luminous, which is clearer in the case of $R_p=0$ (cf.~the left panels in Figs.~\ref{fig.BZFast} and \ref{fig.BZSlow}), due to smaller $|{\bf v}_p|$ around the jet axis, which relaxes the relativistic beaming to the anti-direction to the observer. 

We also confirmed that the results for $0<R_p<40r_g$ present extremely asymmetric jets as inferred from the results for $R_p=0$ and $40r_g$. The large asymmetry is also maintained for larger $R_p$ in our search up to $R_p=100r_g$, which produces a sufficiently wide jet image for M87.

\section{Discussion}\label{sec.discussion}
\subsection{Jet Images in Cold Ideal MHD Treatment}\label{sec.MHD}
In Sec.~\ref{sec.results}, it was found that the disk-threaded model is difficult to produce symmetrically limb-brightened jets in our model and seems to be inappropriate for the M87 jet. We here discuss whether this result changes or not if we give another velocity different from that given by Eq.~(\ref{eq.drift}) as the jet velocity. This is worth considering, since the jet edge part in the disk-threaded model corresponds to the region with $\zeta \le 1$, where the drift velocity, Eq.~(\ref{eq.drift}), may not approach the velocity in cold ideal MHD, which will be the next simplest approximation. Comparing the drift velocity with the velocity in cold ideal MHD jets, we argue expected changes of limb-brightened radio images in the MHD treatment through the modified beaming effects for the observer.

We first focus on the azimuthal speed, since it is critical to the asymmetry in radio images. The toroidal speed in cold ideal MHD outflows under the steady and axisymmetric assumptions is given as follows \citep[e.g.][]{TT13}:
	\begin{equation}\label{eq.vphiMHD}
	\frac{v_\phi}{c} = \frac{1}{\zeta} \left[ 1 -\frac{(1-\tilde{\zeta}^2)\tilde{\Gamma}}{\Gamma}\right],
	\end{equation}
where $\zeta:=R\Omega_\mathrm{F}/c$ and the letters with tilde denote the quantities at the inlet. $\tilde{\Gamma}\sim1$ is the initial Lorentz factor at the inlet. The ratio to the azimuthal speed in our force-free model, the second term in Eq.~(\ref{eq.drift}), is then given by
\begin{equation}\label{vphimhdff_1}
\frac{v_{\phi,\mathrm{MHD}}}{v_{\phi,\mathrm{FF}}} =\frac{g_\mathrm{FF}^2 + \zeta_\mathrm{FF}^2}{g_\mathrm{FF}^2\zeta_\mathrm{FF}\zeta_\mathrm{MHD}}\left(1-\frac{1-\tilde{\zeta}_\mathrm{MHD}^2}{\Gamma_\mathrm{MHD}}\right),
\end{equation}
where the letters with MHD and FF are evaluated in a cold ideal MHD model and our force-free one, respectively. As long as the force-free approximation is reasonable, the shapes of the poloidal magnetic field are the same in ours and MHD.
This assumption of the same-shaped field yields $\zeta_\mathrm{FF}=\zeta_\mathrm{MHD}=\zeta$. Since $g_\mathrm{FF}(\theta, \nu)$ is also determined by the shape of ${\bf B}_p$, we can omit the subscript, FF, hereafter: $g_\mathrm{FF}=g$.

If the ratio given by Eq.~(\ref{vphimhdff_1}) exceeds unity (i.e., $v_{\phi,\mathrm{MHD}}\ge v_{\phi,\mathrm{FF}}$), the disk-threaded model (Case~1) will not be preferred even in cold ideal MHD models due to more asymmetric limb-brightened features (See Appendix~\ref{sec.asymproof} for the proof that faster rotational speeds always lead to more asymmetric images). From Eq.~(\ref{vphimhdff_1}), the inequality $v_{\phi,\mathrm{MHD}}\ge v_{\phi,\mathrm{FF}}$ holds for
\begin{eqnarray}
\Gamma_\mathrm{MHD} &\ge& \left( 1 -\frac{g^2\zeta^2}{g^2 + \zeta^2}\right)^{-1}(1-\tilde{\zeta}^2).\label{eq.sufficient_condition}
\end{eqnarray}
We can put $1-\tilde{\zeta}^2 \sim 1$, since we are now interested in the outer jet with $\zeta \lesssim1$ in the Case~1, which roughly corresponds to $\Psi/A \gtrsim 35.3r_g$ for $|z|<10$~mas (the black curve in Fig.~\ref{fig.fieldKepler}) and $\tilde{\zeta}^2 \lesssim 0.01$. Since $g\sim 1$ at high latitudes where the limb brights (See Fig.~\ref{func_theta_nu}), we can reduce Eq.~(\ref{eq.sufficient_condition}) to
\begin{equation}\label{eq.gamma}
	\Gamma_\mathrm{MHD} \gtrsim 1+\zeta^2.
\end{equation}
That is, if the above condition holds, the simulated radio images for $R_p\ge 40r_g$ would be more asymmetric in the disk-threaded model with cold ideal MHD treatment.

Although the actual value of $\Gamma_\mathrm{MHD}$ in cold ideal MHD treatment could be obtained with a detailed model, it is beyond the scope of this paper. Instead, we here consider whether the limb brightening of the M87 jet can emanate from the region with $\zeta \le 1$ by assuming that the pattern speed observed in the M87 jet corresponds to $\Gamma_\mathrm{MHD}$. \citet{Mertens16} reports that the Lorentz factor of the fast component of the M87 jet exceeds $\sim2$ at $z\gtrsim 3$~mas, which means from Eq.~(\ref{eq.gamma}) that the azimuthal speed should be larger than that in our model provided the limb brightening originates from the region with $\zeta \le 1$. Larger $v_\phi$ enhances the asymmetry of the radio images, which is not consistent with observations.

A possible change of the poloidal speed would always produce problematic jet images: The counter jet becomes more luminous for smaller $|{\bf v}_p|$, while the asymmetry of the emission from `the coming quadrisection' of the jet is enhanced for larger $|{\bf v}_p|$ (See Appendix~\ref{sec.asymproof}).

From the above discussions, the disk-threaded model would not be suitable for the M87 jet even in cold ideal MHD treatment, while MHD numerical simulations should be incorporated for more quantitative discussions. It is noted, on the other hand, that \citet{Mertens16} conjectured a jet launched from a Keplerian accretion disk, based on analyses of observed pattern speeds in the M87 jet with cold ideal MHD treatment. The reason of this discordance should be pursued, although it is beyond the scope of this paper.

\subsection{Effects of the Viewing Angle}\label{sec.discussion.view}
The viewing angle $\Theta$ will be another important parameter as well as $\Omega_\mathrm{F}$ and $R_p$ for producing radio images, since it changes the line-of-sight speed, which strongly beam or debeam the synchrotron emission to the observer. While the viewing angle of the M87 jet is thought to be in the range of $\sim10^\circ \mathrm{-}45^\circ$ based on optical observations of superluminal motion around the HST-1 \citep{Biretta99} and radio observations of proper motion and brightness ratio of the jet and counter jet \citep{Ly07,Hada16,Mertens16}, it will be interesting to study whether the limb-brightened features are kept if the viewing angle were much larger or smaller than the above constraint. We set below $\Theta = 5^\circ$ and $75^\circ$, for example, while the other parameters are the same as in the Case~2 with $a=0.998$ and $R_p = 40r_g$.

\begin{figure*}[t]
	\begin{center}
	\centering	\includegraphics[bb = 0 0 983 582, width=\textwidth
	]{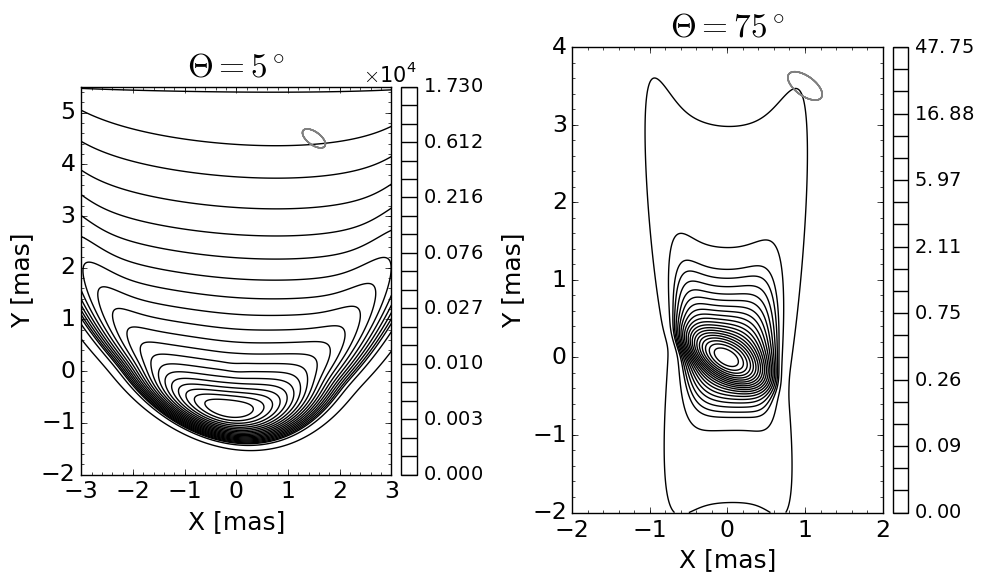}
	\end{center}
	\caption{Same as the right panel in Fig.~\ref{fig.BZFast} but for $\Theta = 5^\circ$ (left) and $75^\circ$ (right), respectively.}
	\label{fig.Theta}
\end{figure*}

The left panel in Fig.~\ref{fig.Theta} presents the result for $\Theta=5^\circ$, which still shows a limb brightening feature. As $\Theta$ decreases, the jet becomes more luminous while the counter jet becomes dimmer due to stronger beaming and debeaming effects to the observer, respectively. At the same time, the jet image is expanded in the transverse direction due to the projection effect.

The right panel in Fig.~\ref{fig.Theta} displays the case for $\Theta = 75^\circ$. The large viewing angle softens the relativistic beaming to the observer. As a result, the jet becomes less luminous while the counter jet becomes more luminous. The limb-brightened feature is still visible in the jet side while it is also apparent in the counter jet.

As presented above, limb-brightened features are observed even for viewing angles much different from our fiducial value. This fact suggests that the limb-brightened features observed in other objects such as Mrk~501, which has the viewing angle of $\Theta\sim5\mathrm{-}15^\circ$ \citep{Giroletti04,Giroletti08}, and Cyg~A, which has $\Theta\sim75^\circ$ \citep{Boccardi16}, are also attributed to the jet structure with magnetic field lines penetrating a fast-spinning BH and non-thermal electrons away from the jet axis. It is interesting, however, that \citet{Boccardi17} came to another conclusion that the jet base of Cyg~A is widely extended and appears to be anchored to the accretion disk. We may need more detailed models with parameters tuned for these objects to make credible conclusions, which will be studied in a forthcoming paper.

\section{Summary \& Conclusions}\label{sec.conclusion}
This paper investigated the relations between the jet structure of AGNs and observed radio images of the jet. We focused on the limb-brightened features observed in some AGNs such as M87 that appears to be largely symmetric to the jet axis. We employed the basically same steady axisymmetric force-free jet model as in \citetalias{BL} but introduced new points of view to produce limb-brightened jets. We compared paraboloidal jets launched from the Keplerian accretion disk and from the central BH. The latter was not investigated in \citetalias{BL}. It was found that they have qualitatively different jet structures, including the jet rotation pattern and speed, which produces qualitatively and quantitatively different radio images even for the same distribution of the emitting particles. We treated the spatial distribution of the non-thermal electrons as a parameter, instead of linking it to some physical processes or just concentrating the particles around the jet axis as in \citetalias{BL}. Simulating radio maps produced by synchrotron radiation, we constrained several important jet parameters for symmetrically limb-brightened jet images.

We demonstrated that symmetrically limb-brightened jets may be launched from a fast-spinning BH with the non-thermal electrons distributed away from the jet axis: We assumed that the magnetic field lines penetrate the BH and the magnetic field lines rigidly rotate with the half angular frequency of the BH. Far away from the jet axis, the jet is sufficiently accelerated to poloidal directions and the jet rotation relatively slows down, which occurs more effectively for larger Kerr parameters. Such a velocity field leads to symmetric jet images with low-luminous counter jets.
Slowly spinning BHs nor the particle distribution concentrated near the jet axis are disfavored: The former results in extremely asymmetric radio emissions due to faster jet rotations while the latter never brightens the edge but ends in a candle-flame-like pattern.

We also suggested that symmetrically limb-brightened jets are not launched from a geometrically thin accretion disk with the Keplerian rotation, which was assumed in \citetalias{BL}. Reasonably, the jet edge is not illuminated unless the non-thermal electrons exist there. The non-thermal electrons away from the jet axis, however, produce strongly asymmetric radio images. This is because the fast jet rotation enhances the difference of the relativistic beaming to the observer between the northern and southern sides of the jet. The luminous counter jet is also a problem of this model in the case of the M87 jet, which is not dimmed because of the slow poloidal speeds in the jet edge. We also pointed out that the disk-threaded model would not be appropriate for the M87 jet even in cold ideal MHD treatment, since the asymmetry of radio images would be enhanced and the counter jet could be more prominent. This challenges the interpretation that the jet is launched from an accretion disk \citep[e.g.][]{Mertens16}.

We cannot exclude, however, the magnetic field lines converging to a narrow ring region on the accretion disk instead of those penetrating the BH horizon, which may cause almost rigidly rotating magnetic field, whereas it should be debatable whether such a concentrated configuration can be realized. It is also noted that the disk-threaded model might relax the asymmetry of jet images and veil the counter jet by assuming an accretion disk rotating with another law that has a weaker dependence on $R$ than for $\Omega_\mathrm{Kep}$ and/or by finely tuning all the parameters in our model, whereas only slowing down the rotation speed is insufficient to solve the problems (See Appendix~\ref{sec.subKepler}). We need more detailed fit to observations in order to totally reject the disk-threaded model.

In our BH-threaded model, the symmetry of radio images is dependent on the Kerr parameter: The symmetric pattern is gradually recovered as the Kerr parameter increases. Therefore, the spin of the central BH could be constrained by fitting the calculated jet image to the observations. Such detailed studies are complimentary to those concentrating directly on the innermost region with upcoming EHT data \citep{Dexter12,Moscibrodzka16}, since the size of observed BH shadows only has a weak dependence on the BH spin \citep[][and references therein]{Psaltis15}. Furthermore, in addition to M87, the limb-brightened jet structures in other AGNs such as Mrk~501 and Cyg~A might be also explained in the same manner with the BH-threaded model. A detailed study for these specific objects will be presented in a forthcoming paper.

It is worth noting again that our results indicate the existence of non-thermal electrons away from the jet axis, which is inevitable to produce limb-brightened images. This constraint is important, since the distribution of the non-thermal particles is one of the most ambiguous points even in more elaborated models using global GRMHD simulations \citep[e.g.][]{Moscibrodzka16}. While the distribution of non-thermal electrons should be given by microscopic processes, our findings might be a hint to search for the site of particle accelerations in relativistic jets. Other sophisticated numerical simulations of relativistic jets, e.g.~\citet{BM10} and \citet{Porth11}, also do not show limb-brightened features because of the assumed spatial distribution of the non-thermal electrons, although their distributions are based on physically motivated models. We also note that \citet{Porth11} assumed jets launched from an accretion disk and, hence, their simulations would not produce a symmetrically limb-brightened jet with a dim counterpart even if they had employed other spatial distributions of emitting particles.

While our simple treatment of relativistic jets lead to suggestive results, a comprehensive treatment with an accretion disk with funnel flows in a more detailed way, e.g., in general relativistic radiation MHD (GRRMHD), must be incorporated in future work, which is inevitable for consistent understanding of the jet-disk system of AGNs. 

\acknowledgements
We thank the participants in the Mizusawa Project Meetings in 2016 and 2017 for fruitful discussions on the M87 jet from various points of view. K.T. and K.T. thank Taiki Ogihara for daily discussions on relativistic jets. The first author thanks Prof. Hiroshi Nagai for his comments on a use of terminology and references. We also thank the anonymous referee for his/her fruitful comments and suggestions. Numerical calculations were performed on Draco, a computer cluster of the Frontier Research Institute for Interdisciplinary Sciences in Tohoku University. This work is partly supported by JSPS Grants-in-Aid for Scientific Research 15H05437 (KT), JP18K03656 (MK), and JP18H03721 (MK, KH), and also a JST grant ``Building of Consortia for the Development of Human Resources in Science and Technology".

\appendix
\section{A. Force-free jet model} \label{sec.model}
\subsection{A.1. Steady Axisymmetric Force-Free Field}\label{sec.ff}
Steady axisymmetric electromagnetic fields have been widely considered in the literature \citep{Mestel61,Okamoto74,BO78, Camenzind86,TT03,VK03,Beskin09}. We review here such fields with the force-free approximation. The basic equations consist of the Maxwell equations and the conservation laws of fluid coupled with electromagnetic field.

We start from the relations that are derived only from the steady axisymmetric condition before imposing the force-free approximation. Analogy to the two-dimensional incompressible flows, a stream function exists for the poloidal magnetic field, by which each magnetic surface is labeled, because of the divergence-free condition of magnetic field in axisymmetric geometry. The poloidal magnetic field, ${\bf B}_p$, is then given as follows \citep{Narayan,BL,TT13}:
\begin{equation}
\label{Bp}
{\bf B}_p = -\frac{1}{R}\frac{\partial \Psi}{\partial z} \hat{\bf R} + \frac{1}{R}\frac{\partial \Psi}{\partial R} \hat{\bf z} =  \frac{1}{r^2\sin \theta} \frac{\partial \Psi}{\partial \theta} \hat{{\bf r}} -\frac{1}{r\sin \theta} \frac{\partial \Psi}{\partial r} \hat{{\bf \theta}} = \frac{1}{R}{\bf \nabla} \Psi \times \hat{\bf \phi},
\end{equation}
where $\Psi := RA_\phi$ is a stream function with $A_\phi$ being the toroidal component of the magnetic vector potential. The vectors with a hat are the unit coordinate bases.
We note that the stream function $\Psi(R,z)$ is essentially the total magnetic flux penetrating within radius $R$ except for a factor of $2\pi$: That is, $\Phi = 2\pi \Psi$ is satisfied for any magnetic flux $\Phi$ \citep{Narayan}.

The steady axisymmetric condition reduces the poloidal component of the Faraday's law, ${\bf \nabla} \times {\bf E} = {\bf 0}$, to the relation that $E_\phi \equiv 0$, where $E_\phi$ stands for  the toroidal component of electric field.  

In the force-free approximation, the plasma inertia and thermal pressure are neglected in dynamics \citep{Narayan,BL}. In this prescription, the fluid contributes only as the charge and current sources. The equation of motion is, hence, reduced to
\begin{equation}
\label{ff}
\rho _e{\bf E} + \frac{1}{c}{\bf j} \times {\bf B} = {\bf 0},
\end{equation}
where $\rho _e$ and ${\bf j}$ are charge and current densities, respectively.

The projection of the both sides of the force-free condition, Eq.~(\ref{ff}), to the direction of ${\bf B}$ yields the condition that the magnetic and electric fields are orthogonal to each other: ${\bf E}\cdot {\bf B} = 0$. Due to the absence of the toroidal electric field, the orthogonal condition gives the electric field as follows \citep{Ly09,TT13}:
\begin{equation}
\label{E}
{\bf E} = -\frac{1}{c}\Omega_\mathrm{F} {\bf \nabla} \Psi = -\frac{R\Omega_\mathrm{F}}{c} \hat{\phi } \times {\bf B},
\end{equation}
where $\Omega_\mathrm{F}(R,z)$ is a scalar function. That is, the surface of $\Psi = \mathrm{const.}$ is also an equipotential surface. Substituting Eq.~(\ref{E}) into the Faraday's law, we obtain the following conservation law from the toroidal component:
\begin{equation}\label{Cons.Omega}
{\bf B}\cdot {\bf \nabla} \Omega_\mathrm{F} = 0,
\end{equation}
which means that $\Omega_\mathrm{F}$ is conserved along a magnetic field line and, hence, is a function of $\Psi$: $\Omega_\mathrm{F} = \Omega_\mathrm{F}(\Psi)$ \citep{TT13}.

The other Maxwell equations recover the corresponding charge and current sources for a given electromagnetic field. The charge density is obtained by the Gauss' law \citep{Narayan}:
\begin{equation}
\rho _e = \frac{1}{4\pi}{\bf \nabla}\cdot {\bf E} = -\frac{\Omega_\mathrm{F}}{4\pi c}\Delta \Psi -\frac{1}{4\pi c}\frac{\diff \Omega_\mathrm{F}}{\diff \Psi} |{\bf \nabla} \Psi|^2,
\end{equation}
while the current density is given by the Amp\`{e}re's law as follows \citep{Narayan}:
\begin{eqnarray}
j_\phi &=& \frac{c}{4\pi}({\bf \nabla}\times {\bf B})_\phi = -\frac{c}{4\pi R}\left(\frac{\partial ^2\Psi}{\partial R^2} - \frac{1}{R}\frac{\partial \Psi}{\partial R} + \frac{\partial ^2 \Psi}{\partial z^2} \right),\\
\label{jp}
{\bf j}_p &=& \frac{c}{4\pi}({\bf \nabla}\times {\bf B})_p = -\frac{c}{4\pi} \frac{\partial B_\phi}{\partial z} \hat{\bf R} + \frac{c}{4\pi R}\frac{\partial (RB_\phi)}{\partial R}\hat{\bf z}.
\end{eqnarray}

The toroidal component of the force-free condition, Eq.~(\ref{ff}), gives a conservation law for the total poloidal current passing trough a toroidal loop of radius $R$, $I \propto RB_\phi$: In fact, the equation gives (${\bf j}\times {\bf B})_\phi =  0$ due to $E_\phi \equiv 0$, which is satisfied only if ${\bf j}_p$ is parallel to ${\bf B}_p$. Comparing these poloidal vectors given by Eqs.~(\ref{Bp}) and (\ref{jp}), one notices that $RB_\phi$ should be a function of $\Psi$ \citep{Narayan}. That is, $RB_\phi$ is conserved along a magnetic field line:
\begin{equation}\label{Cons.Bphi}
{\bf B}\cdot{\bf \nabla}(RB_\phi) = 0.
\end{equation}

We already projected Eq.~(\ref{ff}) to the directions of ${\bf B}$ and $\hat{\bf \phi}$. Because of the orthogonal relations: ${\bf E}\cdot {\bf B}={\bf E}\cdot \hat{\bf \phi}=0$, the projection onto ${\bf E}$ gives a relation independent of the former ones. The last equation determines $\Psi$ for given $\Omega_\mathrm{F}$ and $B_\phi$ as follows \citep{Narayan}:
\begin{eqnarray}
\label{Psi}
\left[1-\left(\frac{R\Omega_\mathrm{F} }{c}\right)^2\right]\left( \frac{\partial ^2 \Psi}{\partial R^2} + \frac{\partial ^2 \Psi}{\partial z^2} \right)- \left[1+\left(\frac{R\Omega_\mathrm{F}}{c}\right)^2\right] \frac{1}{R}\frac{\partial \Psi}{\partial R}
+ \frac{1}{2} \frac{\diff (RB_\phi)^2}{\diff \Psi} - \frac{R^2\Omega_\mathrm{F}}{c^2}\frac{\diff \Omega_\mathrm{F} }{\diff \Psi} |{\bf \nabla}\Psi|^2 = 0,
\end{eqnarray}
which gives the shape of magnetic field that satisfies the force balance in the trans-field direction.

\subsection{A.2. Magnetic Field}\label{sec.psi}
\begin{figure*}
	\centering \includegraphics[bb = 0 0 719 570, width=8cm]{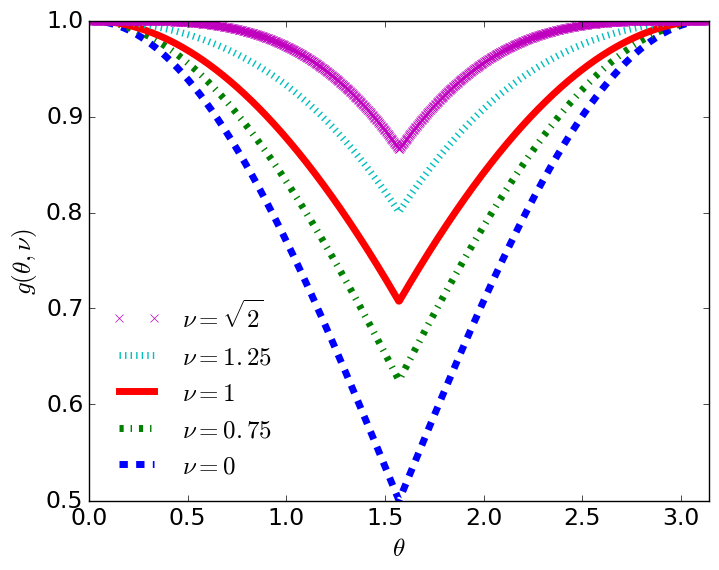}
	\caption{The plots of $g(\theta,\nu)$ for $\nu \le \sqrt{2}$.}
	\label{func_theta_nu}
\end{figure*}

Equation~(\ref{Psi}) becomes singular at the critical surface $R\Omega_\mathrm{F}/c =1$ and a regular solution is found only for an appropriate choice of the functional form of $B_\phi$ for a given $\Omega_\mathrm{F}$. Otherwise, the solution cannot be continuous beyond the singular surface \citep{Fendt97a,CKF99,Beskin09,Takamori14}. Since it is generally a tough task to find such a fully-consistent regular solution, we use a stream function that approximately describes the force-free numerical solution obtained by \citet{TMN08}, which gives a paraboloidal-shaped jet and was also adopted in \citetalias{BL}. The stream function is given by:
\begin{equation}
\label{PsiApprox}
\Psi = Ar^\nu (1\mp \cos \theta),
\end{equation}
where $A$ is a constant that has the dimension of $[r^{2-\nu}B]$ and $\nu$ is the parameter that determines the jet shape. The minus and plus signatures are for $z\ge 0$ and $z<0$, respectively, and the function is symmetric with respect to the equatorial plane, $z=0$.
We note that Eq.~(\ref{PsiApprox}) is a good approximation to the exact solution of the steady axisymmetric force-free field as well as results in numerical simulations \citep{TMN08}. As special cases, Eq.~(\ref{PsiApprox}) gives a split-monopole field for $\nu = 0$ and a paraboloidal field for $\nu = 1$. Since we are interested in collimated jets, we assume $\nu >0$ hereafter. The components of the poloidal magnetic field are given by
\begin{eqnarray}
\label{Br}
B_r &=& \frac{1}{r^2\sin \theta}\frac{\partial \Psi}{\partial \theta} = \pm Ar^{-(2-\nu)} = \pm \frac{\Psi}{R^2}(1\pm \cos \theta), \\
\label{Bth}
B_\theta &=& -\frac{1}{r\sin \theta}\frac{\partial \Psi}{\partial r} = -\nu Ar^{-(2-\nu)}\sqrt{\frac{1\mp \cos \theta}{1\pm \cos \theta}} =  -\nu \frac{\Psi}{R^2}\sin \theta,
\end{eqnarray}
which yield
\begin{equation}
\label{compBp}
B_p =\sqrt{B_r^2 + B_\theta^2}=\frac{2\Psi}{R^2}g(\theta, \nu),
\end{equation}
where
\begin{equation}\label{eq.g}
g(\theta, \nu) = \sqrt{\frac{1\pm \cos \theta}{2}\left[ 1-(1-\nu ^2)\frac{ 1\mp \cos \theta}{2}\right]}.
\end{equation}
We henceforth assume $\nu \le \sqrt{2}$ (for the drift speed less than $c$; See Appendix~\ref{sec.v}). Then, the function $g(\theta, \nu)$ satisfies $\sqrt{1+\nu ^2}/2 \le g(\theta,\nu)\le 1$ as shown in Fig.~\ref{func_theta_nu}. We note that $g(\theta, \nu)$ is reduced to $\cos(\theta/2)$ and $\sin(\theta/2)$ for $z\ge 0$ and $z<0$, respectively, in the case of $\nu=1$.

Corresponding to the given shape of the jet, Eq.~(\ref{PsiApprox}), $B_\phi$ is given by \citep{TMN08}
\begin{eqnarray}
\label{Bphi}
B_\phi = \mp \frac{2\Omega_\mathrm{F} \Psi}{Rc} = \mp \frac{2\Psi}{R^2}\frac{R\Omega_\mathrm{F}}{c}.
\end{eqnarray}
We note that \citetalias{BL} also use the same prescription for $B_\phi$.

The magnitude of the magnetic field is given by Eqs.~(\ref{compBp}) and (\ref{Bphi}) as follows:
\begin{equation}
\label{B}
B=\frac{2\Psi}{R^2} \sqrt{[g(\theta, \nu)]^2 + \left(\frac{R\Omega_\mathrm{F}}{c}\right)^2},
\end{equation}
which gives the following asymptotic relation:
\begin{equation}
\label{asymptB}
B \sim \left\{ 
\begin{array}{cc}
|B_\phi| & \mathrm{for} \ \displaystyle \frac{R|\Omega_\mathrm{F}|}{c}\gg 1 \vspace{1mm}\\
B_p & \mathrm{for} \ \displaystyle \frac{R|\Omega_\mathrm{F}|}{c}\ll 1
\end{array}\right..
\end{equation}

\subsection{A.3. Fluid Velocity}\label{sec.v}
The force-free approximation does not give the fluid velocity, since the fluid inertia is totally neglected and, hence, the motion along a magnetic field cannot be determined. Following \citetalias{BL}, we use the so-called drift velocity as the fluid velocity \citep{Narayan}:
\begin{equation}
\label{v}
{\bf v} = \frac{{\bf E}\times {\bf B}}{B^2} c= -R\Omega_\mathrm{F} \frac{B_\phi}{B^2}{\bf B}_p + R\Omega_\mathrm{F} \frac{B_p^2 }{B^2} \hat{{\bf \phi}}.
\end{equation}
This prescription ensures that (i) the fluid speed does not exceed the speed of light for $\nu \le \sqrt{2}$, (ii) the electric field vanishes in the proper frame, which is consistent with the infinite conductivity, and (iii) the velocity asymptotically approaches the fluid velocity in cold ideal MHD as relativistically accelerated to poloidal directions.

The first and second statements are straightforwardly confirmed by calculation. In fact, the normalized speed of fluid is given by
\begin{equation}
\label{v/c}
\beta := \frac{|\bf{v}|}{c} = \frac{E}{B} = \frac{R|\Omega_\mathrm{F}|}{c}\frac{B_p}{B} = \sqrt{\frac{[g(\theta, \nu)]^2\left(\frac{R\Omega _\mathrm{F}}{c}\right)^2}{[g(\theta, \nu)]^2 + \left(\frac{R\Omega _\mathrm{F}}{c}\right)^2}} \le 1,
\end{equation}
where the equality holds for $g(\theta,\nu)=1$ and $R|\Omega_\mathrm{F}|/c =\infty$. We also note that the azimuthal speed is bound by $c/2$, which can be shown in the same manner.

Equations~(\ref{asymptB}) and (\ref{v}) give the asymptotic relations of the fluid velocity for $R|\Omega_\mathrm{F}|/c \ll 1$ as follows:
\begin{eqnarray}
\beta &\sim & \beta _\phi \sim \frac{R|\Omega_\mathrm{F}|}{c}, \\
\beta_p &\sim& \frac{R|\Omega_\mathrm{F}|}{c}\frac{|B_\phi|}{B_p} = \left( \frac{R\Omega_\mathrm{F}}{c}\right)^2 \frac{1}{g(\theta,\nu)},\\
\Gamma &:= &\frac{1}{\sqrt{1-\beta^2}}  \sim 1+\frac{1}{2}\left( \frac{R\Omega_\mathrm{F}}{c}\right)^2 ,
\end{eqnarray}
where $\beta_p := |{\bf v}_p|/c$ and $\beta _\phi = |v_\phi|/c$ are the normalized poloidal and toroidal speeds, respectively. That is, the fluid velocity is non-relativistic and dominated by the toroidal component. For $R|\Omega_\mathrm{F}|/c \gg 1$, on the other hand, the following relations are obtained:
\begin{eqnarray}
\beta &\sim & \beta _p \sim g(\theta,\nu), \label{eq.asympt_betap}\\
\beta _\phi &\sim&  \frac{R|\Omega_\mathrm{F}|}{c} \frac{B_p^2}{B^2_\phi} = \left( \frac{R|\Omega_\mathrm{F}|}{c}\right)^{-1} [g(\theta,\nu)]^2,\label{eq.asympt_betaphi}\\
\label{GammaAsympt}\Gamma &\sim& \frac{1}{\sqrt{1-[g(\theta,\nu)]^2}}. \label{eq.asympt_gamma}
\end{eqnarray}
That is, the fluid velocity is dominated by the poloidal component, which becomes relativistic as $g(\theta, \nu)$ approaches unity. We note here that, as $g(\theta, \nu) \rightarrow 1$, the leading terms in Eqs.~(\ref{eq.asympt_betap}) and (\ref{eq.asympt_betaphi}) approach those in the asymptotic relations in steady axisymmetric cold outflows in ideal MHD \citep{TT13}:
\begin{eqnarray}
\beta _p &\sim& 1 -\frac{1}{\Gamma^2} -\left(\frac{R\Omega_\mathrm{F}}{c}\right)^{-2} \sim 1, \\
\beta _\phi &=& \left( \frac{R|\Omega_\mathrm{F}|}{c}\right)^{-1}\left[ 1 - \left(1-\frac{\tilde{R}^2\Omega_\mathrm{F}^2}{c^2}\right)\frac{\tilde{\Gamma}}{\Gamma}\right]\sim \left( \frac{R|\Omega_\mathrm{F}|}{c}\right)^{-1},
\end{eqnarray}
which holds for $R|\Omega_\mathrm{F}|/c \gg 1$ and $\Gamma \gg \tilde{\Gamma} \sim 1$, where the letters with tilde denote quantities at the inlet.

\subsection{A.4. Non-thermal Electrons}\label{sec.n}
The number density of the non-thermal electrons, $n$, is assumed to be given by the continuity equation for fluid, ${\bf \nabla} \cdot(n {\bf v})=0$, by following \citetalias{BL}, although it is not so obvious whether the non-thermal electrons obey the equation. For $RB_\phi \Omega_\mathrm{F} \ne 0$, the continuity equation is reduced to 
\begin{equation}
\label{Consn}
{\bf B} \cdot {\bf \nabla} \left( \frac{ n}{B^2} \right) = 0,
\end{equation}
which means that $n$ scales with $B^2$ along a given magnetic field. We also note that the continuity equation also derives the conservations of the ratio of the mass flux to the magnetic flux as in ideal MHD: ${\bf B}\cdot {\bf \nabla}(n|{\bf v_p}|/B_p)=0$ by using Eq.~(\ref{v}). In this paper, we assume the following ring-shaped distribution of the non-thermal electrons on the planes $z=\pm z_1$ $(z_1\ge0)$:
\begin{equation}
\label{Distn}
n(R,\pm z_1) = n_0 \exp\left[ -\frac{(R-R_p)^2}{2\Delta ^2} \right],
\end{equation}
where $R_p$ is the radius where $n$ have the peak on the plane and $\Delta$ gives the width of the ring while $n_0$ is the number density at the peak. We note that \citetalias{BL} considered only $R_p=0$, where the non-thermal electrons are concentrated on the jet axis at $z=\pm z_1$. 

Equations (\ref{Consn}) and (\ref{Distn}) give the number density of the non-thermal electrons at a given point on a magnetic field labeled by $\Psi' $ as follows:
\begin{equation}
\label{n}
n (R,z) = n_0 \frac{B^2(R,z)}{B^2(R_1,z_1)} \exp\left[ -\frac{(R_1-R_p)^2}{2\Delta ^2} \right],
\end{equation}
where $R_1(\Psi ' )$ denotes the radial coordinate of the intersections of $\Psi = \Psi' $ and $z=\pm z_1$. We omit an artificial factor of $(1 -\exp [-r^2/z_1^2])$ in Eq.~(\ref{n}) that was introduced in \citetalias{BL} to reduce plasma in the innermost region $r < z_1$. Our results are not qualitatively different, however, even if the factor is taken into account.

We assume that the distribution of the non-thermal electrons is isotropic in the fluid rest frame and the energy distribution is described by a single power law with an index $p$:
\begin{equation}\label{f}
f(\gamma')  = \left\{
\begin{array}{cc} 
C n' \gamma' {}^{-p} & (\gamma ' _\mathrm{min} \le \gamma' \le \gamma '_\mathrm{max}) \\
0 & \mathrm{otherwise}
\end{array}\right. ,
\end{equation}
where and hereafter quantities with a prime are evaluated in the fluid rest frame. $\gamma'$ is the Lorentz factor of an electron and $\gamma '_\mathrm{min}$ and $\gamma '_\mathrm{max}$ are the minimal and maximal Lorentz factors, respectively. $C$ is a normalization constant, which is given for $p \ne 1$ by \citep{Shibata}
\begin{equation}
C = \frac{(p-1)\gamma '_\mathrm{min}{}^{p-1}}{\displaystyle 4\pi \left[ 1 - \left( \frac{\gamma '_\mathrm{min}}{\gamma '
		_\mathrm{max}}\right)^{p-1}\right] } .
\end{equation}
We assume that the energy distribution is given by Eq.~(\ref{f}) in the entire region, i.e., we assume some energy supplier that compensates the energy loss due to cooling processes such as the synchrotron cooling.

\subsection{A.5. Synchrotron Emissivity in the Fluid Rest Frame}\label{sec.j}

Since we consider highly-relativistic electrons, the synchrotron emission are highly beamed into the direction of the electron motion. In this case, the synchrotron emissivity in the fluid rest frame, $j'_{\omega '}  ({\bf n}')$, is given by \citep{RL,Shibata}
\begin{equation}\label{jd}
j'_{\omega '}({\bf n}') =\frac{\sqrt{3}e^3C n' B'\sin [\psi '({\bf n}')]}{2\pi m_e c^2(p+1)}\left( \frac{m_ec\omega'}{3eB'\sin[\psi '({\bf n}')]}\right)^{- (p-1)/2}\bar{\Gamma} \left( \frac{p}{4} + \frac{19}{12} \right) \bar{\Gamma} \left( \frac{p}{4} - \frac{1}{12} \right),
\end{equation}
where $e$, $m_e$, and $\bar{\Gamma}(\cdots)$ are the elementary charge, the mass of electron, and the gamma function, respectively. $\psi '({\bf n}')$ is the pitch angle of the electrons that directs toward the observer, which are most responsible for producing radio images because of relativistic beaming effects \citep{Shibata}:
\begin{equation}
\cos [\psi ' ({\bf n}')]=  \frac{{\bf n}' \cdot {\bf B}'}{|{\bf B}'|} = \frac{1}{\Gamma (1-\beta \mu)}\frac{{\bf n}\cdot {\bf B}}{|{\bf B}|}.
\end{equation}
In the derivation of Eq.~(\ref{jd}), we used the approximation that $\gamma '_\mathrm{min}$ and $\gamma '_\mathrm{max}$ are sufficiently small and large, respectively, to evaluate an energy integral \citep{RL}. In this case, the energy cutoffs affect the synchrotron emissivity only through the normalization constant $C$.

\section{B. Parameter dependence}\label{sec.dependence}
\subsection{B.1. Fast-spinning BH-threaded models}
We study the dependence of radio images on the parameters that are fixed in the main text. It is important to note that our conclusions in the main text are not changed even if these parameters are altered whereas the radio images are slightly modified. We use the fast-spinning BH-threaded model (Case~2 with $a=0.998$) with $R_p=40r_g$ shown in the right panel in Fig.~\ref{fig.BZFast} as a fiducial model, since it resembles the observed images better than the other models. We change four parameters, $\Delta$ (ring width), $\nu$ (jet shape), $p$ (power index of the energy distribution of electrons), and $M_\mathrm{BH}$ (BH mass) around the fiducial model as in Table~\ref{tab.models} while fixing the other parameters such as $\Omega_\mathrm{F}$ and $a$ as well as $R_p$. Comparing the produced radio images, we discuss the effects of each parameter below.

\begin{figure*}
	\centering \includegraphics[bb = 0 0 1068 422, width=\textwidth
		]{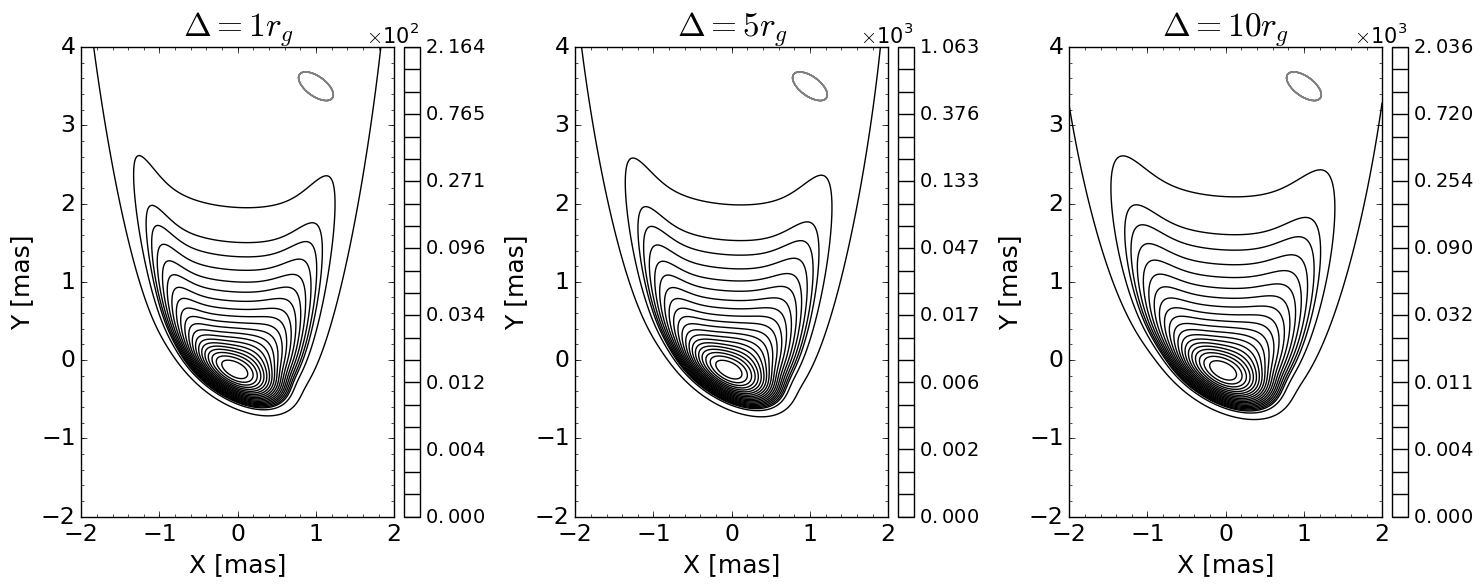}
	\caption{Same as Fig.~\ref{fig.BZFast} but for the models B, A, and C from left to right, respectively.}
	\label{fig.Compare_Delta}
\end{figure*}

The dependence on $\Delta$ is displayed in Fig.~\ref{fig.Compare_Delta}. As naturally expected, the larger $\Delta$ makes radio images wider in the north-south direction, since the electrons are more distributed to the edge region, although the effect is rather limited within this range of $\Delta$. 

\begin{figure*}
	\centering \includegraphics[bb = 0 0 1068 422, width=\textwidth
		]{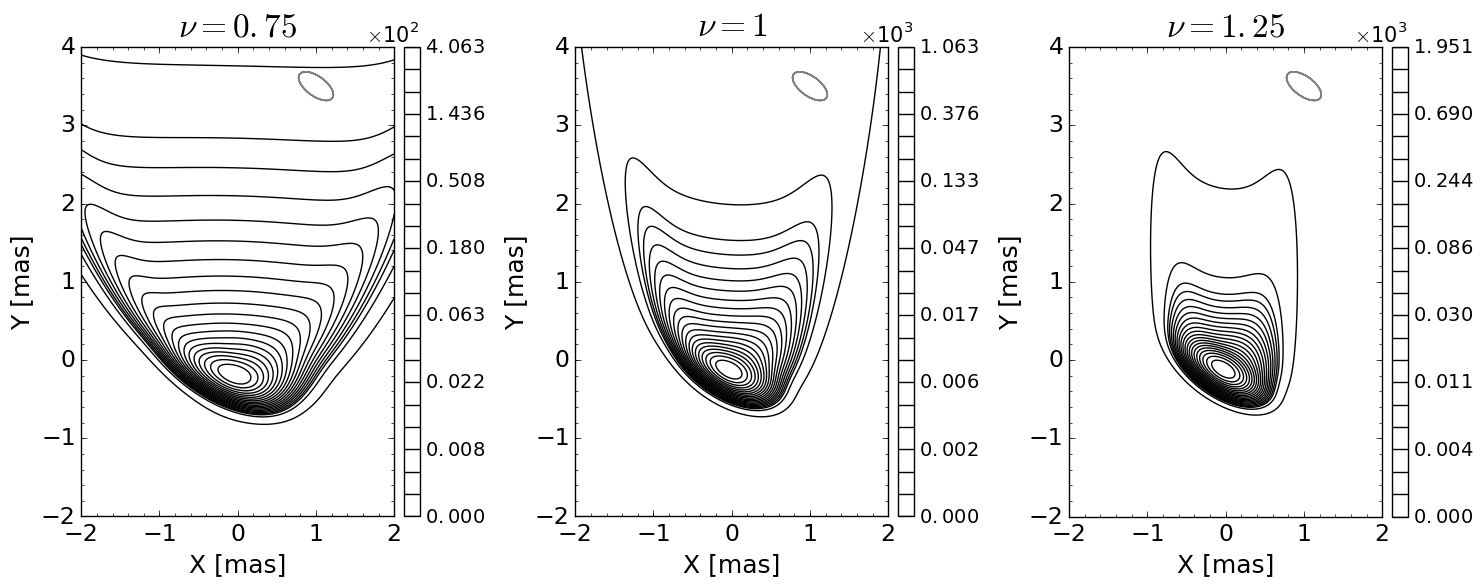}
	\caption{Same as Fig.~\ref{fig.Compare_Delta} but for the models D, A, and E from left to right, respectively.}
	\label{fig.Compare_nu}
\end{figure*}

The dependence on the jet shape is shown in Fig.~\ref{fig.Compare_nu}, where the jet is less (more) collimated in the left (right) panel. We note that the jet shape is expressed by $R\propto z^\xi$ far from the BH ($\theta \ll 1$), where $\xi $ is defined by $\nu = 2 -2\xi$ \citep{TMN08}. That is, $\nu = 0.75$, $1$, and $1.25$ (i.e., $\xi = 0.625$, $0.5$, and $0.375$) give the asymptotic jet shape of $z\propto R^{8/5}$, $R^{2}$, $R^{8/3}$, respectively. The jet shape is clearly reflected to the radio image as tightly collimated jets produce narrower radio images.

\begin{figure*}
	\centering \includegraphics[bb = 0 0 1068 421, width=\textwidth
		]{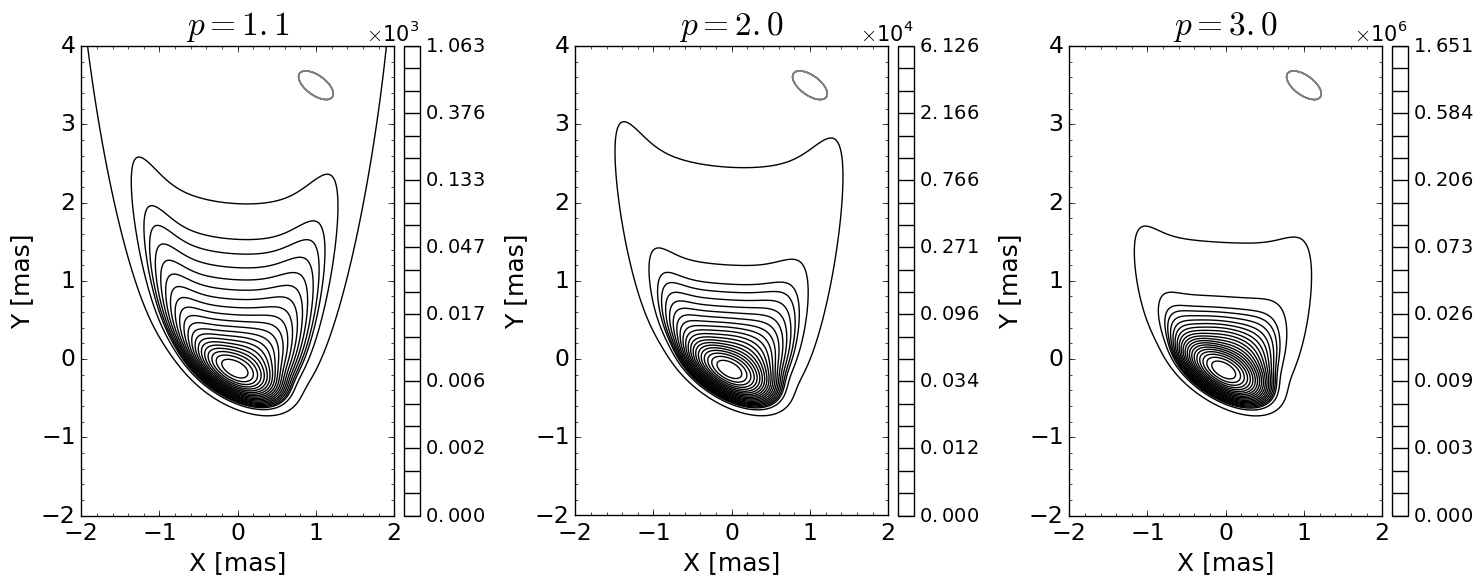}
	\caption{Same as Fig.~\ref{fig.Compare_Delta} but for the models A, F, and G from left to right, respectively.}
	\label{fig.Compare_p}
\end{figure*}

Figure~\ref{fig.Compare_p} manifests that the harder energy distribution of electrons leads to more compact radio images. That is, the contrast of intensity is enhanced for larger $p$, since the difference of the magnetic and velocity fields at different locations is enhanced by $(p-1)/2$ as given in Eq.~(\ref{jd}). The limb-brightened feature becomes discreet, as a result, for large $p$ while it is still discernible in Fig.~\ref{fig.Compare_p}. We note that \citet{Hada16} reported $p\sim2.2\mathrm{-}2.6$ for the M87 jet.

\begin{figure*}
	\centering \includegraphics[bb = 0 0 708 420, width=12cm
		]{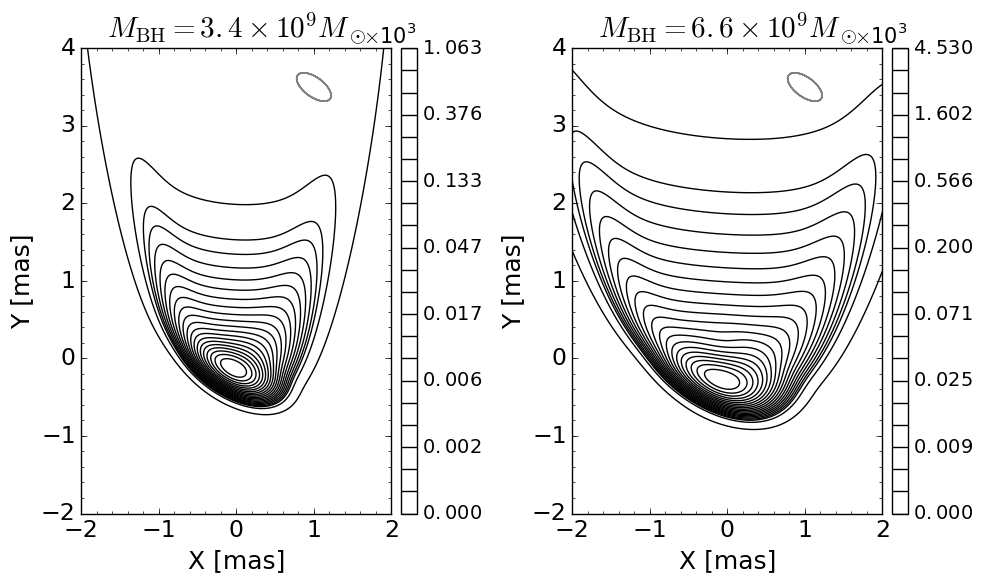}
	\caption{Same as Fig.~\ref{fig.Compare_Delta} but for the models A (left) and H (right). Note that 1~mas corresponds to $\sim 250r_g$ (i.e., $1r_g \sim 4\times10^{-3}$~mas) for the left panel while 1~mas $\sim 130 r_g$ (i.e., $1r_g \sim 8\times10^{-3}$~mas) for the right one.}
	\label{fig.Compare_M}
\end{figure*}

Massive BHs produce ``larger'' radio images as shown in Fig.~\ref{fig.Compare_M}. It should be careful to interpret this result, since the Schwarzschild radius changes as $r_g\propto M_\mathrm{BH}$ while we used $R_p = 40 r_g$ in both models. That is, the electrons are distributed more far away from the jet axis in the model~H with a more massive BH, which directly makes the radio image wider in the $X$-direction. We also note that the BH mass changes $\Omega_\mathrm{F}$, which is proportional to $M_\mathrm{BH}^{-1}$ in the BH-threaded model for a fixed Kerr parameter. Thus, the increase of $M_\mathrm{BH}$ for a fixed $a$ has similar effects as the decrease of $a$ for a fixed $M_\mathrm{BH}$ (Note: $\Omega _\mathrm{F} \propto a/(1+\sqrt{1-a^2})$).

\begin{table}
	\centering
	\caption{Fast-spinning BH-threaded Models}
	\label{tab.models}
	\begin{tabular}{ccccc}
		\hline
		Name & $\Delta$  & $\nu$ & $p$ & $M_\mathrm{BH}$ \\
		& $(r_g)$ & & & $(10^9M_\odot)$ \\
		\hline
		A (fiducial)& $5$ & $1$ & $1.1$ & $3.4$ \\
		B & $1$ & $1$ & $1.1$ & $3.4$ \\
		C & $10$ & $1$ & $1.1$ & $3.4$ \\
		D & $5$ & $0.75$ & $1.1$ & $3.4$ \\
		E & $5$ & $1.25$ & $1.1$ & $3.4$ \\
		F & $5$ & $1$ & $2$ & $3.4$ \\
		G & $5$ & $1$ & $3$ & $3.4$ \\
		H & $5$ & $1$ & $1.1$ & $6.6$ \\
		\hline	\end{tabular}
\end{table}

\subsection{B.2. Sub-Keplerian disk-threaded models}\label{sec.subKepler}
\begin{figure*}
	\centering \includegraphics[bb = 0 0 1068 421, width=\textwidth
	]{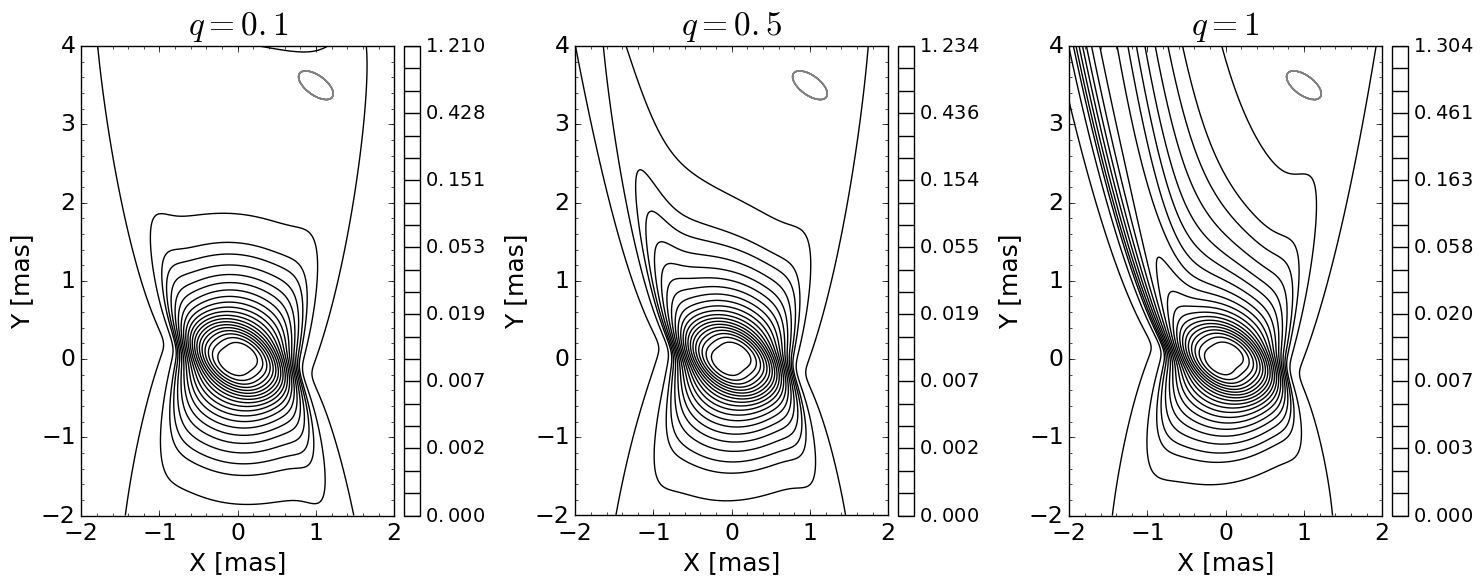}
	\caption{Same as the right panel in Fig.~\ref{fig.Kepler} but for $q=0.1$, $0.5$, and $1$ from left to right, respectively.}
	\label{fig.subKepler}
\end{figure*}

We study here the parameter dependence of radio intensity maps of the disk-threaded model (Case~1). We focus on the disk rotation, which characterizes disk-threaded models, and consider sub-Keplerian motion. Introducing a factor $q$ ($0<q\le1$), we modify Eq.~(\ref{Kepler}) as follows:
\begin{equation}
\label{subKepler}
\Omega_\mathrm{F} =\left\{
\begin{array}{lc}
q\Omega_\mathrm{Kep}(\tilde{R})& (\tilde{R}>R_\mathrm{ISCO}) \\
q\Omega_\mathrm{Kep}(R_\mathrm{ISCO})& (\tilde{R}\le R_\mathrm{ISCO})
\end{array}\right.,
\end{equation}
where $0<q<1$ gives a sub-Keplerian disk while $q=1$ coincides with the Case~1. We pick up the cases with $q=0.1$ and $q=0.5$ for example, while keeping the other parameters the same as in the Case~1. The former is an extreme case of slowly rotating disks and the latter corresponds to ADAFs. Figure~\ref{fig.subKepler} shows the radio intensity maps for $q=0.1$, $0.5$, and $1$ for a reference. As the disk rotation slows down, the radio image recovers the symmetry. However, the limb feature becomes less prominent and the counter jet keeps the brightness, which are inconsistent with observations of M87.

These changes of the radio image are explained as follows: As the disk rotational speed decreases, the light `cylinder' surfaces, both of the curved and vertical ones, shrink inward. The shrink of the light cylinder decreases $v_\phi$ on $\Psi/A\sim35.3r_g$, where most of the emitting particles exist (black lines in Fig.~\ref{fig.fieldKepler}), since $v_\phi$ peaks around the light `cylinder' and decreases toward the jet edge part (See the upper-right panel in Fig.~\ref{fig.fieldKepler}). This is the reason why the asymmetry of radio images is weakened for smaller $q$. The shrink of the light cylinder, at the same time, slows down the poloidal speed, since $|\bf{v}_p|$ becomes smaller apart from the curved light-cylinder surface. Thus, the light emanating from $\Psi/A\sim35.3r_g$ is less beamed as $q$ decreases. As a result, the counter jet keeps the feature and the limb becomes less prominent with respect to the central core, which weakens the limb-brightening feature.

\section{C. The asymptotic shape of the light cylinder in the Case 1}\label{sec.LC}
\begin{figure*}
	\centering \includegraphics[bb = 0 0 470 419, width=8cm]{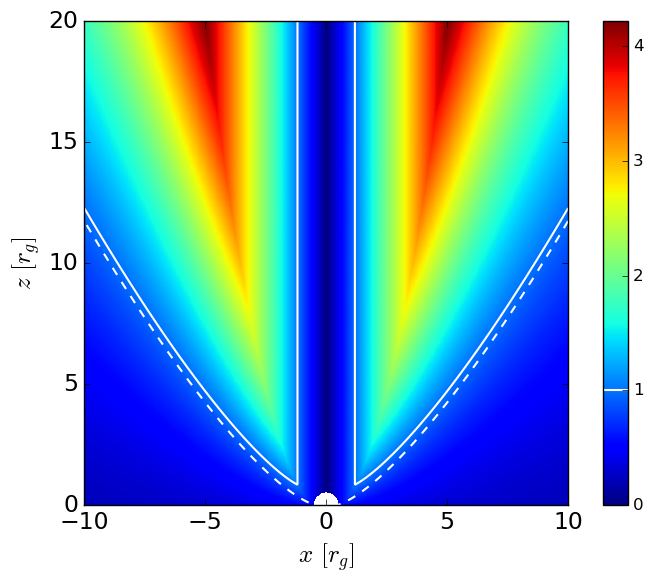}
	\caption{The close-up of the left panel in Fig.~\ref{fig.fieldKepler} around the origin. The newly drawn dashed curve shows the line given by Eq.~(\ref{formula}) (cf. the white solid lines).}
	\label{fig.closeup}
\end{figure*}

We derive here the asymptotic shape of the light `cylinder' in our disk-threaded model (Case~1), which has a curved surface as shown in Fig.~\ref{fig.fieldKepler}. We consider the far zone where each magnetic field line is anchored to the accretion disk far from the gravitational radius, i.e., $\tilde{R} \gg r_G$. Thus, each magnetic field line rotates with $\Omega_\mathrm{F} = \Omega_\mathrm{Kep}(\tilde{R}) \sim \sqrt{GM_\mathrm{BH}/\tilde{R}^3}$. The condition for the light cylinder, $R\Omega_\mathrm{F}/c=1$, is then reduced to the following cubic equation for $z$:
	\begin{equation}
	z^3 + \frac{3R^2}{4}z \mp \frac{R^2(R^2-r_G^2)}{8r_G} = 0, \label{cubic}
	\end{equation}
where we used the relation $\tilde{R} = \Psi/A = \sqrt{R^2+z^2}\mp z$. The real root of Eq.~(\ref{cubic}) is given by
\begin{eqnarray}
z &=&\pm \frac{R}{2}\left[\left({\frac{R}{r_G}}\right)^{1/3} - \left(\frac{R}{r_G}\right)^{-1/3}\right], \label{formula}\\
&\propto&\pm R^{4/3}\ (\mathrm{for}\ R \gg r_G).
\end{eqnarray}
We note that the deviation of the surface given by the above asymptotic relation, Eq.~(\ref{formula}), from the curved surface of the light cylinder is rather small even at $R\gtrsim r_G$ as shown in Fig.~\ref{fig.closeup}.

\section{D. The ratio of the beaming factors on the left and right sides of the jet}\label{sec.asymproof}
In our model, asymmetric radio images are produced mainly due to the difference of the beaming factors, $\delta := 1/[\Gamma (1-\beta\mu)]$, on the left and right sides of the jet with respect to the observer, which is caused by the jet rotation. We here discuss how the asymmetric feature changes when the jet speed changes while the other quantities remain the same values by studying the dependence of the ratio of $\delta$ between the two sides of the jet. We note that the following discussion is quite general and uses only the assumption of an axisymmetric flow.

\begin{figure*}
	\centering \includegraphics[bb = 0 0 609 580, width=8cm]{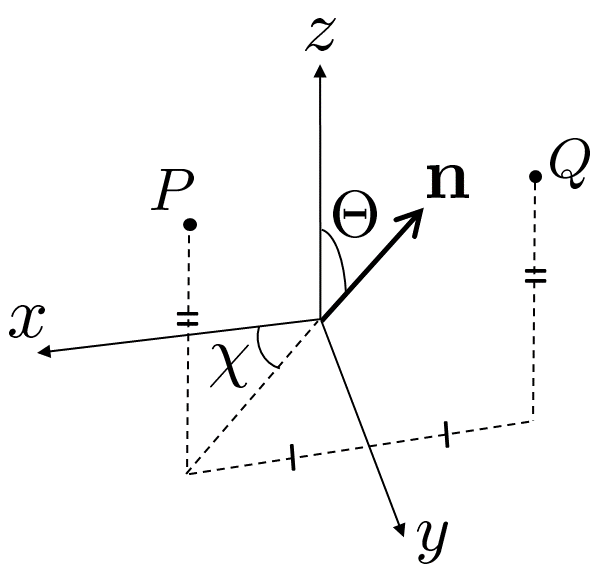}
	\caption{Considered two points, $P$ and $Q$, which are on symmetric positions with respect to the $yz$-plane.}
	\label{fig.coordinates}
\end{figure*}

We use Cartesian coordinates $(x,y,z)$ where the flow is axisymmetric around the $z$-axis and the observer direction is given by ${\bf n} =(0, \sin \Theta, \cos \Theta)$ with $0\le \Theta \le \pi/2$ being the viewing angle. We consider two points $P=(\cos \chi, \sin \chi, z)$ and $Q=(-\cos \chi, \sin \chi, z)$, where $-\pi/2<\chi<\pi/2$ is the azimuthal angle of $P$ measured from the $x$-axis. Note that $P$ and $Q$ are symmetric positions with respective to the $yz$-plane (See Fig.~\ref{fig.coordinates}). Let ${\bf \beta}_P$ and ${\bf \beta} _Q$ be the velocities normalized by $c$ at $P$ and $Q$, respectively. Due to the axisymmetry, they are generally given by
\begin{eqnarray}
	{\bf \beta} _P = (\beta _R \cos \chi - \beta_\phi \sin \chi,  \beta_R\sin \chi + \beta _\phi \cos \chi, \beta _z), \label{betaP}\\
	{\bf \beta} _Q = (-\beta _R \cos \chi - \beta_\phi \sin \chi,  \beta_R\sin \chi - \beta _\phi \cos \chi, \beta _z),\label{betaQ}
\end{eqnarray}
where $\beta _R$ and $\beta _\phi$ are, respectively, the radial and azimuthal velocities at $P$ (or equivalently at $Q$). We can assume $\beta _\phi \ge 0$ without loss of generality. Equations~(\ref{betaP}) and (\ref{betaQ}) yield
\begin{eqnarray}
\mu _P := {\bf \beta}_{P} \cdot {\bf n} = \beta _R \sin \chi \sin \Theta + \beta _\phi \cos \chi \sin \Theta + \beta _z \cos \Theta, \\
\mu _Q := {\bf \beta}_Q \cdot {\bf n} = \beta _R \sin \chi \sin \Theta - \beta _\phi \cos \chi \sin \Theta + \beta _z \cos \Theta.
\end{eqnarray}
The ratio of the beaming factors at $P$ and $Q$ is given by $\varepsilon := \delta _P/\delta _Q = (1-\beta \mu _Q)/(1-\beta \mu _P)$, where $\beta = |{\bf \beta} _P|= |{\bf \beta} _Q|$. Rotations with $\beta _\phi > 0$ lead to $\delta > 1$, which means that light emitted from $P$ is more beamed to the observer than from $Q$. Without rotations ($\beta _\phi = 0$) or viewed from the $z$-axis ($\Theta =0$), $P$ and $Q$ become equivalent for the observer and, hence, $\varepsilon$ is unity.

We first note that faster rotations always enhance the difference of the relativistic beaming to the observer between $P$ and $Q$ unless $\Theta = 0$. That is, the ratio $\varepsilon$ monotonically increases with $\beta _\phi$, which follows from 
\begin{eqnarray}
\frac{\partial \varepsilon}{\partial \beta_\phi} = \frac{2(\alpha + \beta _\phi ^2) \cos \chi \sin \Theta}{\beta (1-\beta \mu _P)^2} \ge 0,
\end{eqnarray}
where the inequality holds, since $\alpha :=\beta ^2[1-\beta (\beta _R \sin \chi \sin \Theta + \beta _z \cos \Theta)]$ turns out to be non-negative as follows:
\begin{eqnarray}
\alpha 
&\ge & \beta ^2[1-\beta (\pm \beta _R \sin \Theta + \beta _z \cos \Theta)],\\
&=& \beta ^2[1-\beta \beta _p\sin(\Theta + \varphi _\pm)],\\
&\ge& 0,
\end{eqnarray}
where the plus and minus sings are for $\beta_R > 0$ (i.e., expanding flows) and $\beta _R \le 0$ (converging flows), respectively. $\beta_p = \sqrt{\beta _R^2 + \beta _z^2}$ is the poloidal speed and $\varphi_\pm$ is given by $\cos \varphi_\pm =\pm \beta_R/\beta _p$ and $\sin \varphi_\pm =\beta_z/\beta _p$.

The behavior of the ratio $\varepsilon$ is more complicated to the change of the poloidal velocities, $\beta _R$ and $\beta _z$, as shown below. The differential of $\varepsilon$ with respect to $\beta _R$ is given by
\begin{eqnarray}
\frac{\partial \varepsilon}{\partial \beta_R} &=&\frac{2\beta _\phi (\beta _R + \beta^3\sin \chi \sin \Theta) \cos \chi \sin \Theta}{\beta (1-\beta \mu _P)^2}.
\end{eqnarray}
The signature of $\partial \varepsilon/\partial \beta_R$ depends on the signature of $\beta _R + \beta^3\sin \chi \sin \Theta$, which can be either of positive or negative. It should be noted, however, that $\partial \varepsilon/\partial \beta_R$ is non-negative for $\chi >0$ and $\beta _R > 0$. That is, the increase of the radial speed amplifies the difference of the beaming effects between $P$ and $Q$ in the half side of the expanding flow that is near to the observer when divided by the $xz$-plane, unless $\Theta =0$ nor $\beta_\phi =0$. This is the case for our jet model.

The differential of $\varepsilon$ with respect to $\beta _z$ is given by
\begin{eqnarray}
\frac{\partial \varepsilon}{\partial \beta_z} &=& \frac{2\beta _\phi (\beta _z + \beta^3\cos\Theta) \cos \chi \sin \Theta}{\beta (1-\beta \mu _P)^2},
\end{eqnarray}
which can be positive or negative, depending on the signature of $\beta _z + \beta^3\cos\Theta$. It is, however, worth noting that $\partial \varepsilon/\partial \beta_z$ is non-negative for $\beta _z >0$, i.e., when the outflow comes toward the observer. If applied to our jet model, it means that the increase of $\beta _z$ enhances the difference of the beaming effects between the left and right sides of the jet whereas it is not always the case for the counter jet.

\end{document}